\documentclass[twocolumn,showpacs,showkeys,preprintnumbers,amsmath,amssymb]{revtex4}
\usepackage{graphicx}% Include figure files
\usepackage{dcolumn}% Align table columns on decimal point
\usepackage{bm}% bold math

\begin{document}
%\preprint{}

\title{POLARIZATION OF BREMSSTRAHLUNG AT ELECTRON SCATTERING\\IN ANISOTROPIC MEDIUM}% Force line breaks with \\

\author{M.V. Bondarenco}
\email{bon@kipt.kharkov.ua}
\affiliation{%
Kharkov Institute of Physics and Technology, 1 Academic St., 61108
Kharkov, Ukraine.
}%

\date{\today}% It is always \today, today,
             %  but any date may be explicitly specified

\begin{abstract}
Bremsstrahlung from relativistic electrons is considered under
conditions when some transverse direction of momentum transfer is
statistically preferred. It is shown that in the dipole
approximation  all the medium anisotropy effects can be accumulated
into a special modulus-bound transverse vector, $\bm N$. To
exemplify a target with $N^2\sim1$, we calculate radiation from
electron incident at a small angle on an atomic row in oriented
crystal. Radiation intensity and polarization dependence on the
emission angle and frequency for constant $\bm N$ is investigated.
Net polarization for the angle-integral cross-section is evaluated,
which appears to be proportional to $N^2/2$, and decreases with the
increase of the photon energy fraction. A prominent feature of the
radiation angular distribution is the existence of an angle at which
the radiation may be completely polarized, in spite of the target
complete or partial isotropy -- that owes to existence of an
origin-centered tangential circle for polarization in the fully
differential radiation probability kernel. Possibilities for
utilizing various properties of the polarized bremsstrahlung flux
for preparation of polarized photon beams and for probing intrinsic
anisotropy of the medium are analyzed.
\end{abstract}

\pacs{61.85.+p, 41.60.-m, 78.70.-g, 95.30.Gv}% PACS, the Physics and Astronomy
                             % Classification Scheme.
\keywords{azimuthally anisotropic scattering, polarized radiation, equivalent photon approximation, stereographic projection}%Use showkeys class option if keyword
                              %display desired
\maketitle

\section{Introduction}

Relativistic electrons interacting with matter are efficient sources
of gamma-radiation, which may be applied either for probing nuclei
and hadrons \cite{Proc gamma-nucl}, or to deliver information about
the medium the electrons are passing through. The full set of the
radiation characteristics includes photon polarization, which
correlates with the preferential direction of acceleration of the
radiating particle in the medium, as well as with the photon
emission azimuthal angle. Detection techniques sensitive to
$\gamma$-quantum polarization have been developed to date in a
rather wide range of photon energies \cite{RHESSI,nucl
polarimeters,pol-through-or-cryst}.

For electron scattering on one atom, which is practically a
spherically-symmetric object, the whole problem is axially
symmetric, and thereby net polarization of the radiation (i. e.,
when integrated over the relativistically small emission angles)
vanishes. In contrast, in condensed matter, particularly in
crystals, due to correlation of atomic positions, the aggregate
fields can be highly anisotropic. But in what concerns
bremsstrahlung, it is essential to recall that the major
contribution to the radiation intensity comes from spatial regions
with highest electromagnetic field strength, whereas for atoms those
regions are perinuclear and non-overlapping, containing
centrally-symmetric fields, anyway. To compete with this
contribution, soft action of the atoms on the radiating fast
electron must be enhanced, in coherent manner. So far, examples of
highly azimuthally anisotropic motion or scattering of electrons in
crystals were basically restricted to (planar) channeling
\cite{Gemmel-Kumakhov}, and coherent bremsstrahlung
\cite{Diambrini,Ter-Mik}. These cases demand perfect crystals, high
initial electron beam collimation, and precise crystal orientation
with respect to the incident beam. An extra benefit is the fair
monochromaticity of the emitted $\gamma$-radiation.

But in case if one is interested in radiation polarization only,
regardless of its monochromaticity, and so seeks only scattering
azimuthal anisotropy, not periodicity of the electron motion in the
medium, it seems sufficient to get by with a much rougher
experimental setup. Taking a sufficiently thin crystal cut about
perpendicularly to one of the main crystallographic directions, one
can expect atomic chains along this direction to maintain their
orientation within the crystal thickness. An elementary interaction
of a fast charged particle with a string making a relatively small
angle with the particle direction of motion already introduces an
asymmetry between two transverse directions for particle deflection;
the ordering of strings in transverse directions on the crystal area
is not prerequisite.

The purpose of the present article is to calculate scattering
azimuthal asymmetry and the bremsstrahlung polarization for the
abovementioned physical problem of electron-string interaction, and
estimate minimal conditions for the crystal quality and orientation,
beam collimation degree, etc. Thereat, it may not suffice to deal
with scattering on one string, since a statistical ensemble of
strings contributes. Besides that, even for thin crystals the
thickness may be large enough for failure of factorization between
scattering and radiation, so that radiation and motion in the
external field become an inseparable problem.

Concerning prospects of statistical and non-factorized description
of the radiation spectral intensity, a simplifying property of
electron propagation in atomic matter is that small (relative to the
electron mass) momentum transfers to atoms dominate \footnote{That
is true for elastic scattering, but the latter, in fact, dominates
over inelastic when nuclear charges are $Z\gg1$, as coherent
contribution vs. incoherent ($Z^2$ vs. $Z$).}, justifying equivalent
photon \cite{WW,Bertulani,Ter-Mik}, also known as dipole
approximation \cite{BKStrakh}. The value of the latter approximation
is that it makes the radiation differential probability simply a
quadratic form in the transferred momentum. That permits statistical
averaging over the momentum transfers in matter, basically, in a
model-independent way. As we will show, all the anisotropy effects
get absorbed into a single transverse vector, pointing along the
preferential direction of momentum transfer in the medium, and
having the absolute value related to the asymmetry degree. However,
at a substantial non-factorizability of radiation and scattering,
the magnitude of this vector can depend on the emitted photon
energy.

The plan of the paper is as follows. In Sec.~\ref{sec:fact} we
define the equivalent photon approximation for the bremsstrahlung
process, for simplicity initially presuming the scattering
factorization. We discuss the kinematical relations obeyed by the
polarized photons. In Sec.~\ref{sec:integr} we turn to averaging
over momentum transfers in matter which requires relaxing the
scattering factorization assumption, and analyze possibilities for
obtaining high azimuthal anisotropy with macroscopic targets. In
Sec.~\ref{subsubsec:quant} we specialize to the problem of electron
interaction with atomic strings, first evaluating bremsstrahlung one
one string, and then estimating the effects of multiple scattering.
In Sec.~\ref{sec:ang} we evaluate spectra and angular distributions
for the polarization bremsstrahlung yield at an arbitrary, but (for
simplicity) photon momentum independent, macroscopic anisotropy
parameter. A summary is given in Sec.~\ref{sec:summary}.

\section{Basic bremsstrahlung properties (factorization conditions)}\label{sec:fact}

The general statement of bremsstrahlung problem assumes a
relativistic electron (mass $m$, initial 4-momentum $p=(E,\bm p)$)
scattering on a static solid target, not necessarily intrinsically
isotropic, and detecting $\gamma$-quanta in the typical direction
close to $\bm p$, hence most probably emitted by incident fast
electrons, and most probably not more than one $\gamma$-quantum per
electron. In a fully exclusive event, when the final electron has a
well-defined 4-momentum, as well (denote it as $p'=(E',\bm p')$),
the 4-momentum conservation law and the mass shell conditions read
\footnote{We set $c=1$ and further will put  $\hbar=1$.}:
\[
   p=p'+k+q,
\]
\[
   p^2=p'^2=m^2, \quad k^2=0,
\]
where $k=(\omega,\bm k)$ is the emitted $\gamma$-quantum momentum,
and $q=(0,\bm q)$ -- the total momentum transferred from the
electron to the target.

Polarizations of the photons exchanged with the target can be
regarded as certain, described by a vector $e^\mu$, granted that the
target is heavy and recoilless. To view it as a source of a static
potential in the laboratory frame, we assume
\begin{equation}\label{e,q-def}
    e=(1,\bm{0})
\end{equation}
or Lorentz invariantly,
\begin{equation}\label{stat-gauge}
    e\cdot q=0,\qquad e^2=1.
\end{equation}
The final photon polarization vector $e'^\mu$ in any gauge satisfies
\begin{equation*}
    e'\cdot k=0,\qquad e'^2=-1.
\end{equation*}
The natural gauge for the final real photon is, in the lab frame
\begin{equation*}
    e'=\left(0,\bm e'\right).
\end{equation*}
Initial and final electron bispinors $u$, $u'$ obey Dirac equations
\begin{equation}\label{Dirac}
    (p\cdot\gamma-m)u=0,
\qquad
    \bar u'(p'\cdot\gamma-m)=0,
\end{equation}
and the normalization conditions
\[
\bar u u=\bar u' u'=2m.
\]

Large 4-momenta $p$, $p'$, $k$ in the lab frame are nearly
collinear. Their spatial direction we will let to be $Oz$; spatial
vector components orthogonal to $Oz$ with be marked with a subscript
$\perp$. The naturally emerging ratios of the large collinear
momenta will defined in terms of their energy components:
\begin{equation}\label{xs}
    \frac\omega E=x_\omega\lesssim1,\qquad \frac{E'}E=1-x_\omega\sim1.
\end{equation}

\subsection{Scattering factorization conditions}
\begin{figure}
\includegraphics[width=85mm]{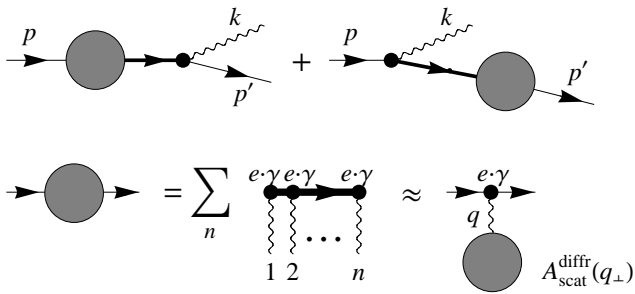}
\caption{\label{fig:Feynm} Factorization of the photon radiation
amplitude at electron scattering on a static source of electric
field. Line thickness reflects the electron virtuality, or fastness
of the process: thinnest lines represent real electrons, medium
thickness lines -- propagation between the scattering and photon
emission, thick line -- propagation between the scatterings.}
\end{figure}

Let us begin with a simplified problem of radiation under the
scattering factorization condition. The scattering factorization
property implies finite-range interaction during an
ultra-relativistic collision, when the time of the scattering is
much shorter than the typical time of decay processes, including the
radiation emission (a sort of impulse approximation). That makes the
photon predominantly emitted from the electron `legs' prior to and
after the scattering. To remind how it formally manifests itself in
different popular frameworks, first refer to the target rest frame,
where one observes relativistic extension (by Lorentz-factor
$\gamma$) of the radiation formation length \cite{Ter-Mik}
\begin{equation}\label{lform}
    l_{\mathrm{form}}=q_z^{-1}\sim \frac{\gamma(1-x_\omega)}{x_\omega m}
\end{equation}
($q_z$ stands for the longitudinal component of typical $q$ in the
process, more precisely -- see Eq.~(\ref{qz}) below), relative to
the field localization domain $\sim r_\text{a}$ (the atomic radius).
Thus, the factorization condition is
\begin{equation}\label{qzra<<1}
    r_\text{a}\ll l_{\mathrm{form}},\quad \mathrm{i.\,e.,}\quad q_zr_\text{a}\ll 1.
\end{equation}
From another viewpoint, in a frame where the electron is
non-ultra-relativistic and evolves together with its electromagnetic
proper-field at times $\sim m^{-1}$, the target atom becomes
longitudinally Lorentz-contracted to the size $\sim
r_{\text{a}}/\gamma$, and appears to the radiating electron as a
short kicker, leading again to same condition (\ref{qzra<<1}).
Finally, when working in the momentum representation, say, in terms
of Feynman diagrams, the emitted real photon typically changes the
electron virtuality (square of its 4-momentum in a virtual state) by
amount $\sim m^2$. As for momentum exchange with the target,
individual longitudinal transfers $q_z^{(i)}$ which are of the order
of $r_{\text{a}}^{-1}$ (though $\sum_iq_z^{(i)}=q_z$ is
kinematically restricted to be $\ll r_\text{a}^{-1}$) make
denominator of the electron's propagator relativistically large --
but a proper compensation arrives from the energy numerator, typical
for vector coupling theories (the same reason as for finiteness of
forward cross-sections -- see, e. g., \cite{Cheng-Wu}). However, if
the real photon is emitted in between the momentum exchanges with
the target, it splits a hard electron propagator into two hard ones,
without a numerator compensation. Therefore, largest are
contributions from diagrams in which the real photon is the first or
the last one in the sequence, leading to the same ordering of
radiation and scattering as inferred from the previous spatial
consideration -- see Fig.~\ref{fig:Feynm}. The technical profit from
the encountered ordering is that it allows factorizing the amplitude
of the entire process into a (nearly on-shell) amplitude of
scattering and the amplitude of radiation at a single scattering act
\cite{Olsen-Maximon-Wergeland}.

As for the Dirac matrix structure of the scattering amplitude, for
small-angle scattering it is particularly simple. In each
contribution to the amplitude from propagation between the
scatterings (say, on the initial end)
\[
e\cdot\gamma\frac{p\cdot\gamma-\sum q^{(i)}\cdot\gamma+m}{(p-\sum
q^{(i)})^2-m^2}e\cdot\gamma,
\]
the spin numerator can always be recast as
\begin{eqnarray}
% \nonumber to remove numbering (before each equation)
  e\cdot\gamma\left(p\cdot\gamma-\sum
q^{(i)}\cdot\gamma+m\right)e\cdot\gamma\qquad\qquad\nonumber\\
  =2e\cdot p
e\cdot\gamma+\left(-p\cdot\gamma+\sum
q^{(i)}\cdot\gamma+m\right)e^2.\label{numerator}
\end{eqnarray}
With $e\cdot p/m=\gamma$ (Lorentz-factor), $\bar u'e\cdot\gamma
u/\bar u'u\sim\gamma$, the second term in (\ref{numerator}) is
generally $\mathcal{O}\left(\gamma^{-2}\right) $ relative to the
first one and can be neglected within the accuracy of the
factorization approximation (\ref{qzra<<1}). Proceeding so in all
orders, the matrix scattering amplitude can be written as
$e\cdot\gamma A_\text{scat}^\text{diffr}(q_\perp)$, where
$A_\text{scat}^\text{diffr}(q_\perp)$ is the \emph{spin-independent}
near-forward angle scattering amplitude including all orders in
perturbation theory. Physically, it can be regarded as diffractive
(potential, non-absorptive), whereby its spin independence looks
intuitive.

Ultimately, the factorization theorem for the small-angle
bremsstrahlung process assumes the form
\begin{equation}\label{fact-amp}
T_\text{fact}=A^\text{diffr}_\text{scat}\left(q_{\perp}\right)\sqrt{4\pi}e
M_\text{rad}\left(q_{\perp},k\right)\left\{1+\mathcal
O\left(q_zr_{\text{a}}\right)\right\},
\end{equation}
with
\begin{eqnarray}\label{Mrad}
M_\text{rad}=\bar{u}' \bigg( \frac{e'^*\cdot\gamma(p\cdot\gamma-
q\cdot\gamma+m)e\cdot\gamma}{2p'\cdot k}\qquad\quad\nonumber\\
- \frac{e\cdot\gamma(p'\cdot\gamma+
q\cdot\gamma+m)e'^*\cdot\gamma}{2p\cdot k} \bigg) u,
\end{eqnarray}
the tree-level radiation matrix element, and $
A^\text{diffr}_\text{scat}(q_{\perp})$ -- the exact elastic
scattering amplitude abridged of the conserved electron bispinors.
If we normalize $A^\text{diffr}_\text{scat}\left(q_{\perp}\right)$,
in accord with its diffractive interpretation, so that the
diffractive scattering differential cross-section expresses as
\begin{equation}\label{Anorm}
d\sigma^\text{diffr}_\text{scat}=\left|
A^\text{diffr}_\text{scat}\left(\bm q_\perp\right)\right|^2\frac{
d^2q_{\perp} }{(2\pi)^2},
\end{equation}
the factorization theorem for probabilities will read
\begin{eqnarray}
  d\sigma_\text{rad}&=&\frac{1}{2E}\left|T_{fi}\right|^2\frac{d^2q_{\perp}}{(2\pi)^2 2E'}\frac{d^{3}k}{(2\pi)^{3} 2\omega}\nonumber\\
  &=&d\sigma^\text{diffr}_\text{scat}\left(\bm
q_\perp\right)dW_\text{rad}\left(\bm
q_\perp,k\right).\label{cross-sect-fact}
\end{eqnarray}
Here,
\begin{equation}\label{Wraddef}
dW_\text{rad}=\frac{4\pi
\alpha}{4EE'}\left|M_\text{rad}\right|^2d\Gamma_{k}
\end{equation}
is the differential probability of single photon emission into a
Lorentz-invariant phase space volume
\begin{equation}\label{dGamma def}
d\Gamma_{k}=\frac{d^3k}{(2\pi)^{3} 2\omega}.
\end{equation}

It is important to note that the formulated theorem does not require
the softness of the emitted photon in the sense that its energy may
be of the order of initial electron energy. That is why in
Eq.~(\ref{Mrad}) the spin structure of the radiation matrix
amplitude is essential.

\subsection{Comptonization conditions}
Although  $q^2$ is not subject to an exact mass-shell restriction,
but in atomic matter $q$ is typically soft, i. e.,
\begin{equation}\label{cond}
   \left|q^2\right|\sim r_{\text{a}}^{-2}\ll m^2.
\end{equation}
Other kinematic invariants in the problem, $p\cdot q$ and $p'\cdot
q$, are $\sim m^2$, if $x_\omega\sim1$. So, everywhere except in the
overall factor to be isolated later on, $q^2$ can be neglected, thus
leading to the equivalent photon approximation:
\begin{equation}\label{Compt approx}
    M_\text{rad}\approx M_\text{rad}\big|_{q^2=0}=M_\text{Compt}.
\end{equation}
The initial, equivalent photon polarization $e$ is real.

We will not be interested in electron polarization effects herein.
Averaging of $|M|^2$ over initial electron's and summation over
final electron's polarizations is simplified in photon gauge of
orthogonality to the initial electron momentum:
\[
e=e_p,\quad e'=e'_p,
\]
\begin{equation}\label{gauge}
e_p\cdot p\overset{\text{def}}=0\approx e_p\cdot q,\quad e'_p\cdot
p\overset{\text{def}}=0=e'_p\cdot k,
\end{equation}
and leads to the result
\begin{eqnarray}\label{KN2}
\left\langle \left| M_{{\text{Compt}}} \right|^2
\right\rangle_{\text{el.spin}} =2\left(\left|e_p\cdot
e'_p\right|^2+\left|e^*_p\cdot
e'_p\right|^2\right)\qquad\qquad\,\nonumber\\
+\left|e_p\right|^2\! \frac{(q\cdot k)^2}{p\cdot q p\cdot
k}+\!\left(\left|e_p\cdot e'_p\right|^2\!-\!\left|e^*_p\cdot
e'_p\right|^2\right)\!\left(\frac{p\cdot q}{p\cdot k}+\frac{p\cdot
k}{p\cdot
q}\right)\nonumber\\
=4(e_p\cdot e'_p)^2+e_p^2 \frac{(q\cdot k)^2}{p\cdot q p\cdot
k},\qquad\qquad\qquad\qquad\qquad\,\,
\end{eqnarray}
where in the last line we took into account that for equivalent
photons $e^*=e$. Therefore, the final photon polarization $e'$ will
be linear, too. In the given gauge, it appears that the final photon
polarization correlates only with $e_p$, but not with the particle
momenta.

In case of truly real photons, when $e_{p}^2=-1$, Eq.~(\ref{KN2})
turns to the Klein-Nishina's formula for unpolarized electrons and
linearly polarized initial and final photons \cite{Klein,BLP}, but
for pseudo-photons the polarization vector square significantly
differs from 1 (see Eq.~(\ref{e2}) below). To apply Eq.~(\ref{KN2})
in arbitrary gauge (in particular, to bremsstrahlung in the
laboratory frame, where the initial electron is relativistic, and
$e$ and $e'$ in physically motivated gauges are by far not
orthogonal to $p$), it suffices to substitute for $e_{p}$, $e'_{p}$
their gauge-invariant representations
\begin{subequations}\label{17}
\begin{eqnarray}
% \nonumber to remove numbering (before each equation)
  e_p=e-q\frac{e\cdot p}{p\cdot q},\,\,\label{ep}\\
  e'_p=e'-k\frac{e'\cdot p}{p\cdot k}.\label{e'p}
\end{eqnarray}
\end{subequations}

To determine the approximation accuracy, begin with noting that
among kinematic invariants, we were neglecting $q^2$ compared to
$p\cdot k\sim x_\omega m^2$. This is a source of relative errors
$1+\mathcal{O}\left(\frac{q^2}{x_\omega m^2}\right)$. But entire
radiation amplitude is of order $e_p\cdot e'_p\approx E\frac{\bm
p'\cdot\bm e'}{p'\cdot k}-E'\frac{\bm p\cdot\bm e'}{p\cdot
k}\sim\frac{q_z}{q_\perp}\sim\gamma\frac{q_\perp }{x_\omega m}$,
compared to which we neglect contributions like $E'\frac{\bm
p\cdot\bm e'}{(p\cdot k)^2}q^2\sim \gamma\frac{q^2}{x^2_\omega
m^2}$. Thereby, the dipole approximation relative accuracy is not
better than
\begin{equation}\label{accur}
    1+\mathcal{O}\left(\frac{q_\perp}{x_\omega m}\right)
\end{equation}

Accuracy (\ref{accur}) implies the condition
\begin{equation}\label{}
    x_\omega \gg
    \frac{q_\perp}m\sim\frac1{mr_{\mathrm{a}}}\sim\alpha\sim
    10^{-2}\quad \mathrm{(dipole \, approximation)}.
\end{equation}
Besides that, factorization condition (\ref{qzra<<1}) implies
\begin{equation}\label{hard bound}
   \gamma\gg\frac{x_\omega}{1-x_\omega}\frac{1}{\alpha}\gg1\quad \mathrm{(factorization\, on\, one\,
   atom)}.
\end{equation}
But at $\gamma>10^2$ the necessary conditions are fulfilled
comfortably enough, allowing for $x_\omega$ variation virtually the
whole interval from 0 to 1.

%It will be instructive to note that the first term of (\ref{KN2}),
%proportional to square of (\ref{A5}), equals to ... for a spinless
%charged particle, whereas the remainder may be interpreted as being
%due to electron spin effects, herein averaged over (`spin light').
%For this term the accuracy of the dipole approximation ( ) is
%typically (when $x_\omega\sim1$, $k_\perp\sim m\gg q$) worse than
%for a spinless particle.

It should be minded, of course, that when folded with the
differential cross-section of scattering in fields with Coulombic
cores, the accuracy of the equivalent photon approximation is known
to diminish down to logarithmic \cite{WW,BLP}. But as mentioned in
the Introduction, we will be seeking ways to overcome this.

In what follows, generally we will not be indicating the
approximation accuracy explicitly.

\subsection{Differential probability of polarized bremsstrahlung in the lab
frame. Angular distribution at a definite $\bm q$}\label{subsec:lab}

In the ultra-relativistic kinematics, more appropriate variables
describing the emitted photon are $x_\omega$ defined by (\ref{xs})
and the \emph{rescaled} angle of emission with respect to initial
(or final) electron momentum:
\begin{eqnarray}\label{Theta def}
    \bm\Theta&=&\frac Em\left(\frac{\bm k}\omega-\frac{\bm
    p}E\right)\equiv \frac{E'}m\left(\frac{\bm k}\omega-\frac{\bm
    p'}{E'}\right)-\frac{\bm q}m\nonumber\\
    &\approx&\frac{E'}m\left(\frac{\bm k}\omega-\frac{\bm
    p'}{E'}\right)
\end{eqnarray}
(in the dipole approximation, when $q/m\ll\Theta\sim1$, initial
electron and final electron and photon momenta lie approximately in
the same plane). In their terms, denominators of Eqs.~(\ref{KN2},
\ref{17}) can be presented as
\begin{equation}\label{dip-denom}
    p\cdot k= E'q_z, \qquad p\cdot q\approx Eq_z,
\end{equation}
where
\begin{equation}\label{qz}
    q_z=\frac{mx_{\omega}\left(
    1+\bm{\Theta}^2\right)}{2\gamma(1-x_{\omega})}.
\end{equation}
The kinematic ratio entering Eq.~(\ref{KN2}):
\begin{equation}\label{pq-pq`}
\frac{(q\cdot k)^2}{p\cdot q p\cdot
k}\approx\frac{x_\omega^2}{1-x_\omega}.
\end{equation}
The equivalent photon polarization vector square (in product with
$\left|A_\text{scat}(\bm{q}_{\perp})\right|^2$ representing the
equivalent photon flux) is
\begin{equation}\label{e2}
-e^2_p=-1-q^2\left(\frac{e\cdot p}{p\cdot
q}\right)^2\simeq-1+\frac{\bm q^2}{q_z^2}=\frac{\bm
q^2_\perp}{q_z^2}\gg1,
\end{equation}
and the photon polarization correlator reduces to
\begin{eqnarray}\label{ee}
e_p\cdot e'_{p}&=&-E\frac{e'\cdot q}{p\cdot q}+\left(q\frac{e\cdot
p}{p\cdot q}-e\right)\cdot k\frac{e'\cdot p}{p\cdot k}\nonumber\\
&\equiv&-\frac{\bm q \cdot \bm e'}{q_z}\nonumber\\
&+&\frac{\omega E}{E'q_z}\left(q\frac{e\cdot p}{p\cdot
q}-e\right)\cdot\left(\frac{k}\omega-\frac
pE\right)e'\cdot\left(\frac pE-\frac{k}\omega\right)\nonumber\\
&=&-\frac1{q_z}q_{\perp i}G_{im}e'_m,
\end{eqnarray}
where
\begin{equation}\label{Gim}
G_{im}(\bm\Theta) = \delta_{im}-\frac2{1+\Theta^2}\Theta_i\Theta_m.
\end{equation}
The final photon phase space element simplifies to
\begin{equation}\label{dGamma small-angle}
d\Gamma_{k}=\frac{d\omega}{2\omega}\frac{\omega^2do_{\bm
k}}{(2\pi)^3} \approx\frac{
dx_\omega}{x_\omega}\frac{m^2x_\omega^2}{16\pi^3}d^2\Theta.
\end{equation}

Inserting all the ingredients (\ref{pq-pq`}-\ref{ee}, \ref{dGamma
small-angle}) into Eq.~(\ref{KN2}), and this latter to
Eq.~(\ref{Wraddef}), one arrives at the final expression for the
bremsstrahlung differential probability \footnote{Inspection of
Eqs.~(\ref{Wdip}, \ref{unpol}) shows that besides the overall
proportionality to $\bm{q}_{\perp}^2$, both
$dW_\text{dip}/d\Gamma_{k}$ and $dW_\text{unpol}/d\Gamma_{k}$, are
still dependent on $\hat{\bm q}_\perp$, and thus on $\bm{q}_\perp$
as a whole -- in spite of the condition $\bm{q}_{\perp}^2\ll m^2$.
Effects of residual azimuthal correlations in equivalent
photon-induced reactions  are, in principle, known, at least, for
some other problems \cite{Budnev}. This is not a genuine
factorization failure, since physical conditions of the latter hold
well, but rather a modification due to the polarization carried by
the equivalent photon flow. (The author thanks I.~F.~Ginzburg for
communication on this point). It is also true that the present
effect disappears when $dW_\text{dip}/d\Gamma_{k}$ is integrated
over the azimuthal directions of $\bm\Theta$ and summed over
$\bm{e}'$ (as is usually done in application to inclusive peripheral
particle production \cite{Bertulani}, or to energy losses of fast
charged particles in matter, being the original concern of Fermi,
Weizs\"{a}cker, and Williams \cite{WW}).}:
\begin{eqnarray}\label{Wdip}
    x_\omega\frac{dW_{\text{dip}}}{dx_\omega d^2\Theta}=\frac{m^2x_\omega^2}{16\pi^3}\frac{4\pi
\alpha}{4EE'}\left\langle \left( M_{{\text{Compt}}} \right)^2
\right\rangle_{\text{el.spin}}\qquad\quad\nonumber\\
=\frac{\alpha}{4\pi^2}\frac{\bm{q}_{\perp}^2}{m^2\left(
1+\Theta^2\right)^2} \left\{4(1-x_{\omega})\left(G_{im}\hat
q_{m\perp}e'_i
\right)^2+x_\omega^2\right\},\nonumber\\
\end{eqnarray}
where
\[
\hat{\bm
q}_\perp=\frac{\bm{q}_{\perp}}{\left|\bm{q}_{\perp}\right|}.
\]

One may notice that in the limit $x_\omega\to0$ intensity
(\ref{Wdip}) reduces to that of classical particle dipole radiation
in an undulator \cite{undul}. In fact, vector $\textbf{ a}_1$ of
\cite{undul} is similar to our vector $\mathsf G\hat{\bm q}_\perp$.
%\footnote{In Eq.~(4.16) of \cite{undul} $\sin^2\varphi$, presumably, must read as $\sin2\varphi$.}
Although the undulator motion is of permanently accelerated type,
not scattering, the description in those cases is largely similar,
because Fourier transform expands small-angle deflections at
scattering in periodic modes, anyway.

On the other hand, to establish relation of notation (\ref{Wdip})
with familiar notations of bremsstrahlung theory, one may pass to
the dipole approximation in the semi-classical radiation amplitude
\begin{eqnarray}
\!\!\!\!\!\bm I_\text{semi-cl}\cdot\bm{e}'&=&\left(\frac{\bm
p'}{p'\cdot k}-\frac{\bm
p}{p\cdot k}\right)\cdot\bm{e}'\label{Isemi-cl full}\\
&\equiv&\left(\frac{\bm{v}'}{\omega-{\bm k}\cdot\bm v'}-\frac{\bm
v}{\omega-{\bm k}\cdot\bm v}\right)\cdot\bm{e}',
\end{eqnarray}
with
\begin{equation}\label{vel}
    \bm v={\bm p}/E,\qquad \bm v'={\bm p'}/{E'}
\end{equation}
being the initial and final ultra-relativistic electron velocities
(nearly unit vectors, $1-\bm v^2=1-\bm v'^2=\gamma^{-2}\ll1$).
Inserting here $\bm v'=\bm v+\bm{\chi}$ and expanding up to linear
terms in the small electron deflection angle $\bm\chi$ gives
\begin{equation}\label{Isem-dip}
\bm
I_\text{semi-cl}\cdot\bm{e}'\approx\frac1\omega\left\{\frac{2\bm\chi}{\gamma^{-2}+\bm\theta^2}
-\frac{4\bm\theta\cdot\bm\chi}{\left(\gamma^{-2}+\bm\theta^2\right)^2}\bm
\theta\right\}\cdot\bm{e}',
\end{equation}
where
\begin{equation*}\label{}
    \bm\theta=\bm k/\omega-\bm v=\bm\Theta/\gamma
\end{equation*}
is the radiation angle. This corresponds to the infra-red-leading
term of Eq.~(\ref{Wdip}). Notation (\ref{Isem-dip}, \ref{Wdip})
compared to (\ref{Isemi-cl full}) has the merit of not involving
large cancelations, and manifestly exposes the polarization
direction -- pointing along vector $\mathsf G \hat{\bm q}_\perp$.

\begin{figure}
\includegraphics{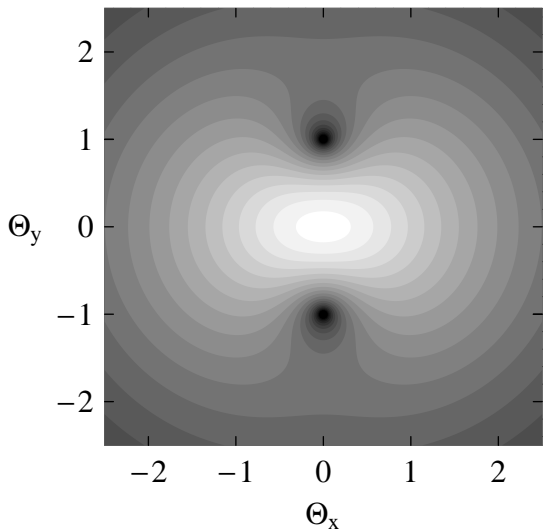}
\caption{\label{fig:UnpolInt} Logarithm of unpolarized radiation
intensity (\ref{unpol}), at $x_\omega\to0$, as a function of
$\bm\Theta$ (the radiation angle vector in units of $\gamma^{-1}$).
The direction of $\bm q_\perp$ is chosen to be along $y$-axis. A
pair of dips (black spots) is manifest. With the increase of
$x_\omega$ the dips get filled in.}
\end{figure}

The unpolarized probability corresponding to Eq.~(\ref{Wdip}) is
obtained by summing it over the independent directions of $\bm{e}'$:
\begin{eqnarray}\label{unpol}
x_\omega\frac{dW_{\text{unpol}}}{dx_\omega d^2\Theta}=\sum_{\bm{e}'}
x_\omega\frac{dW_{\text{dip}}}{dx_\omega d^2\Theta}
\qquad\qquad\qquad\qquad\qquad\nonumber\\
=\frac{\alpha}{2\pi^2}\frac{\bm{q}_{\perp}^2}{m^2\left( 1+
\Theta^2\right)^2} \left\{2(1-x_{\omega})\left(\mathsf G \hat{\bm
q}_\perp\right)^2 +x_\omega^2\right\}.\quad
\end{eqnarray}
Two representations for $(\mathsf G \hat{\bm q}_\perp)^2$ are of
utility:
\begin{subequations}
\label{nu^2}
\begin{eqnarray}
(\mathsf G \hat{\bm q}_\perp)^2&=&1-\frac4{\left(1+\Theta^2\right)^2}\left(\bm\Theta\cdot \hat{\bm q}_\perp\right)^2\label{nu^2=1-}\\
&\equiv&\frac{\left(\bm\Theta +\hat{\bm
q}_\perp\right)^2\left(\bm\Theta -\hat{\bm
q}_\perp\right)^2}{\left(1+\Theta^2\right)^2}.\label{nu^2=+-}
\end{eqnarray}
\end{subequations}
Eq.~(\ref{nu^2=1-}) shows that $(\mathsf G \hat{\bm q}_\perp)^2$ has
the upper bound 1, whereas (\ref{nu^2=+-}) proves that it can
decrease to zero:
\begin{equation*}
0\leq(\mathsf G \hat{\bm q}_\perp)^2\leq1.
\end{equation*}
The angular distribution of (\ref{unpol}) is shown in
Fig.~\ref{fig:UnpolInt}.

The fully differential radiation probability in all variables $\bm
q_\perp$, $\bm\Theta$, $\bm e'$, $x_\omega$ is rarely subject to
observation -- usually measurements are more inclusive,
corresponding to integration over all variables but one or two.
Nonetheless, to be able to predict behavior of the integrated
probability, it is pre-requisite to understand the features of the
integrand. Those main features are listed below.

\begin{figure}
\includegraphics{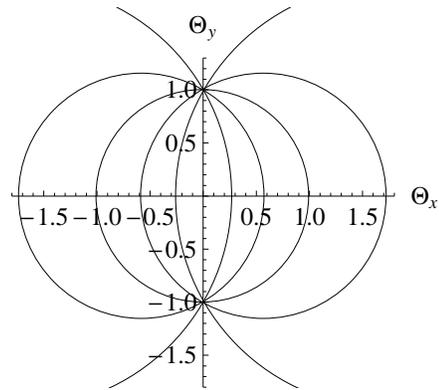}
% Here is how to import EPS art
\caption{\label{fig:Graph1} Photon polarization alignment in the
$\bm\Theta$ plane. The curves tangential to polarization are
mathematical circles. The direction of $\bm q_\perp$ is along
$y$-axis. The polarization degree depends on $x_\omega$.}
\end{figure}

\begin{enumerate}
    \item \textbf{Quasi-Rutherford asymptotics in $\Theta$.} At large
$\Theta$, radiation intensity (\ref{Wdip}) falls off as
$\Theta^{-4}$, i. e. follows essentially the same law as the
Rutherford scattering cross-section. This is a general consequence
of proportionality of the amplitude to one hard propagator -- in the
present case of electron, not of a photon.  In fact, in
Sec.~\ref{sec:spher} we shall yet encounter a kind of `transient
asymptotics' at moderate $\Theta$ (if $x_\omega$ is sufficiently
small).
    \item \textbf{Polarization alignment along circles at a definite $\hat{\bm q}_\perp$.}
It is easy to show by straightforward solution of the ordinary
differential equation
\begin{equation}\label{diff eq}
\frac{d\Theta _y}{d\Theta_x}=\frac{G_{ym}\left(\Theta
_x,\Theta_y\right)q_{\perp m}}{G_{xm}\left(\Theta
_x,\Theta_y\right)q_{\perp m}},
\end{equation}
that curves tangential to the vector field of polarization
directions $\mathsf G \hat{\bm q}_\perp$, are a family of circles
\begin{equation}\label{circles}
    \bm\Theta^2+\mathrm{const}[\bm q\times\bm\Theta]_z=1
\end{equation}
passing through two knot points
\begin{equation}\label{Theta plus-minus}
    \bm\Theta_\pm=\pm\hat{\bm q}_\perp,\qquad \mathrm{(polarization\,knots, \, intensity\, dips)}
\end{equation}
(see Fig.~\ref{fig:Graph1}). Along with $\mathsf G \hat{\bm
q}_\perp\underset{\bm{\Theta}\to\pm\hat{\bm q}_\perp}\approx
\bm{\Theta}\mp \hat{\bm q}_\perp$, in those points to zero drops the
polarization.
    \item \textbf{A pair of intensity dips at a definite $\hat{\bm q}_\perp$.}
As is indicated by Eq.~(\ref{nu^2=+-}), there exists a pair of
$\bm\Theta$ values, specifically (\ref{Theta plus-minus}), at which
$(\mathsf G \hat{\bm q}_\perp)^2$ turns to zero. Those directions
correspond to minima in the radiation intensity at aa definite
$\hat{\bm q}_\perp$ (see Fig.~\ref{fig:UnpolInt}).
\end{enumerate}

Features 2, 3, and coincidence of knots and the dips seem to be non
accidental. Light on their origin will be thrown in the next
subsection.

\subsection{View from the initial electron rest frame. Stereographic
projection}\label{subsec:stereogr}

The features of the polarization angular distribution are realized
best when we take a view from the initial electron rest frame
(IERF).
\begin{equation}\label{p-compon}
    p=(m,\bm{0}).
\end{equation}
In the ultra-relativistic limit $e\cdot p\to \infty$. Vector $e$
tends to be light-like:
\begin{equation}\label{e-compon}
    e=\frac1m(e\cdot p,-e\cdot p,\bm{0}_\perp).
\end{equation}

The final photon momentum in IERF has all components commensurable
\begin{equation}\label{k-compon}
    k=\left(\Omega, K_z,\bm{K}_\perp\right)\quad
    \bm{K}_\perp=\bm{k}_\perp,\quad K_z=\Omega\cos\Psi,
\end{equation}
and $e'_p$ is a spatial vector orthogonal to $\bm{K}$:
\begin{equation}\label{e'-compon}
    e'_p=(0,\bm{e}'_p),\quad \bm{e}'_p\cdot
    \bm{K}=0,\quad |\bm{e}'_p|=1.
\end{equation}

Vector $q$, which must be orthogonal to $e$, belongs to the light
front:
\begin{equation}\label{q-compon}
    q=\left(\frac{p\cdot q}m,-\frac{p\cdot
    q}m,\bm{q}_\perp\right),\quad \frac{p\cdot q}m\sim x_\omega m\quad
    q^2=-\bm{q}^2_\perp,
\end{equation}
and in this frame is not transverse, however, vector $e_p$, with
which the final photon polarization correlates according to
Eq.~(\ref{ep}), is transverse:
\begin{equation}\label{ep-compon}
    e_p=\left(0,0,-\bm{q}_\perp\frac{e\cdot p}{p\cdot
    q}\right)\qquad (e_p\cdot q\neq0).
\end{equation}
So, the polarization vector correlator
\begin{equation}\label{y}
    e_p\cdot e'_p\approx-\frac{\bm q_{\perp}}{q_z}\cdot\bm e'_p
\end{equation}
has the usual dipole appearance analogous to that of
non-relativistic classical electrodynamics. Therewith, polarization
$\bm{e}'_p$ in IERF is distributed along meridians of a sphere of
radiation directions, the polar axis being set by the vector
$\bm{q}_\perp$.

To reproduce Eqs.~(\ref{ee}, \ref{Gim}), it remains to relate $\bm
e'_p$ in IERF with $\bm e'$ in the lab frame. This relation appears
to be particularly simple, too. The consided vectors have equal
moduli $|\bm e'_p|=|\bm e'|=1$, and equal components orthogonal to
the photon scattering plane $(\bm K, Oz)$ (because these components
are not altered neither by the boost along $Oz$, nor the gauge
transformation -- translation along 4-vector $q$). Hence, components
in the plane $(\bm K, Oz)$ must have the same norm and be related by
a pure rotation. Obviously, since $\bm e'$ is nearly orthogonal to
$Oz$, whereas $\bm e'_p$ is orthogonal to $\bm K$, the angle of this
rotation is just the angle $\bm \Psi$ between $Oz$ and $\bm K$:
\begin{equation}\label{}
    \bm e'_p=\mathsf R_{\bm\Psi(\bm\Theta)}\bm e'
\end{equation}
($\mathsf R_{\bm\Psi(\bm\Theta)}$ is a product of an operator of
gauge transformation and of a boost operator). So, one can view
(\ref{y}) as
\begin{equation}\label{}
    e_p\cdot e'_p=-\frac{\bm q_\perp}{q_z}\mathsf G\bm e',
\end{equation}
where
\begin{subequations}
\begin{eqnarray}
    \mathsf G&=&\mathsf P_\perp \mathsf R_{\bm\Psi}\mathsf
    P_\perp\\
    &=&\left(\mathsf P_\perp-\mathsf P_{\bm
    k_\perp}\right)+\cos\Psi\mathsf P_{\bm
    k_\perp}\nonumber\\
    &\equiv&\mathsf P_\perp-\left(1-\cos\Psi\right)\mathsf P_{\bm
    k_\perp}\label{G=PP},
\end{eqnarray}
\end{subequations}
$\mathsf P_\perp$ being an operator of projection onto the plane
$\perp Oz$, and $\mathsf P_{\bm k_\perp}$ -- a projector onto
direction $\hat{\bm k}_\perp$.

Finally, to construct for $\mathsf{G}$ an explicit representation in
terms of $\bm{\Theta}$, one first needs to specify
$\Psi(\bm{\Theta})$. By definition,
\begin{equation}\label{Theta-perp}
    \bm\Theta=\gamma\frac{\bm k_\perp}\omega.
\end{equation}
Here $\omega$, the photon energy in the lab, is related with the
energy and momentum by a light-cone dilation
\begin{equation}\label{Doppler}
    \omega=\gamma\left(\Omega+K_z\right)\qquad (\mathrm{relativistic\, Doppler\, effect\, eqn.})
\end{equation}
Together Eqs.~(\ref{Theta-perp}, \ref{Doppler}) give
\begin{equation}\label{TTheta K}
    \bm\Theta\left(\frac{\bm{K}}\Omega\right)=\frac{\bm
    k_\perp}{\Omega+K_z}\qquad (\mathrm{ligh\, aberration\, formula}).
\end{equation}
\begin{figure}
\includegraphics[height=55mm]{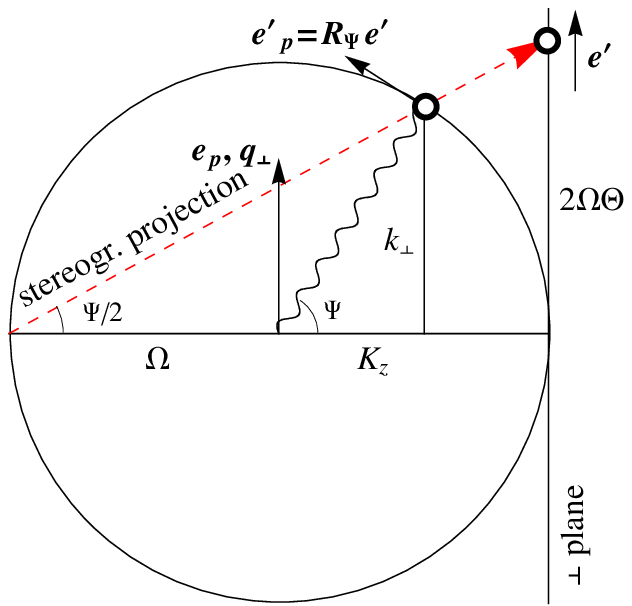}
\caption{\label{fig:Stereograph} Correspondence between the photon
emission angles and polarizations: in the initial electron rest
frame ($\bm\Psi$, $\bm e'_p$) and in the high-energy transverse
plane ($\bm\Theta$, $\bm e'$). For other notations see text. Vectors
shown in bold have also components transverse to the plane of the
figure. The dashed red line designates the stereographic projection
implied by the proportion (\ref{TTheta K}). As a result of
projection of the meridional circles from the sphere onto the plane
one obtains Fig.~\ref{fig:Graph1}.}
\end{figure}
Taking the square of Eq.~(\ref{TTheta K}), one relates $\Theta^2$
with $\cos\Psi$:
\begin{equation}\label{1+cos}
    \Theta^2=\tan^2\frac\Psi2\equiv\frac{1-\cos\Psi}{1+\cos\Psi}.
\end{equation}
Inverting this,
\begin{equation}\label{}
    1-\cos\Psi=\frac{2\Theta^2}{1+\Theta^2},
\end{equation}
and thus the operator entering (\ref{G=PP}) is
\begin{equation}\label{1-cos}
    \left(1-\cos\Psi\right)\mathsf P_{\bm k_\perp}=\frac2{1+\Theta^2}\bm\Theta\otimes\bm\Theta.
\end{equation}
Inserting (\ref{1-cos}) to (\ref{G=PP}), we return to formulas
(\ref{ee}, \ref{Gim}).

The benefit from the described alternative view is clarification of
geometrical origin of the radiation polarization angular
distribution in IERF and in the lab. Relation (\ref{TTheta K}), with
the help of Fig.~\ref{fig:Stereograph}, may be interpreted as a
projection of unit vectors $\bm K/\Omega$, i. e. points on the
radiation direction sphere in IERF, onto tangential plane of photon
emission angles $\bm\Theta$ in the lab, performed from the sphere
point opposite to the plane tangency point. Such a construction is
known in geometry as stereographic projection
\cite{Rosenfeld-Sergeeva}. It has a few remarkable properties.

\begin{description}
  \item[Theorem 1.] Points of the sphere symmetric relative to the
  plane $z=0$ at a stereographic projection pass into points on the
  plane, symmetric relative to the circle $\left|\bm\Theta\right|=1$,
  in the sense that the product of distances from this points to the origin
  equals 1.
\end{description}
The proof is trivial: If $\Psi_1=\frac\pi2-\alpha$,
$\Psi_2=\frac\pi2+\alpha$, then
\begin{equation*}\label{}
    \Theta_1^2\Theta_2^2=\frac{1-\cos\Psi_1}{1+\cos\Psi_1}\frac{1-\cos\Psi_2}{1+\cos\Psi_2}=\frac{1-\sin\alpha}{1+\sin\alpha}\frac{1+\sin\alpha}{1-\sin\alpha}=1.
\end{equation*}

The value of Theorem 1 is that it explains unobvious symmetries in
the ultrarelativistic particle radiation angular distribution as a
manifestation of sufficiently obvious symmetry under Cartesian
inversion in the dipole-radiating particle rest frame.

Another useful property is
\begin{description}
  \item[Theorem 2.] Stereographic projection maps
any circle on the sphere to a circle on the plane.
\end{description}
(The proof thereof is more complicated, and we refer for it to the
literature \cite{Rosenfeld-Sergeeva} and we do not reproduce it
here). Now, since polarization of dipole radiation in IERF is
distributed on the radiation direction sphere along meridional
circles, with the polar axis pointing along $\bm q_\perp$, there is
no wonder that bremsstrahlung polarization tangential curves
evaluated in Eq.~(\ref{circles}) and exhibited in
(Fig.~\ref{fig:Graph1}) are circles, too. (Not surprising either is
the existence of the knot pair, which are just projections of the
knots on the sphere).

The practical value of the circular polarization alignment pattern
will become clear in Sec.~\ref{sec:ang}, where we consider polarized
radiation angular distributions averaged over momentum transfers
$\bm q_\perp$. The dipole radiation angular distribution shape does
not depend on $\left|\bm q_\perp\right|$, whereas  averaging over
$\hat{\bm q}_\perp$ directions implies superimposing polarized
radiation intensity distributions rotated with respect to each
other. This generally suppresses the radiation polarization, except
around angle $\left|\bm\Theta\right|=1$, where polarization is
rotationally invariant. As for intensity minima of dipole radiation
along polar directions in IERF, they make the radiation equatorially
concentrated in IERF, which projects into a bar at
$\left|\bm\Theta\right|<1$ in the lab frame (cf.
Fig.~\ref{fig:UnpolInt}). That bar will, of course, be rounded by
averaging over $\hat{\bm q}_\perp$. The existence of intensity and
polarization maxima can be used for extraction of polarized
radiation beams by the angular collimation technique, as will be
considered in Sec.~\ref{sec:ang}.

\section{Statistical averaging over momentum transfers in matter}\label{sec:integr}

Let us now proceed to description of radiation on a solid target.
Momentum $\bm q_\perp$ imparted to the target is normally beyond
detection and has to be integrated over, with the weight
$(2\pi)^{-2}\left|A^\text{diffr}_\text{scat}\left(\bm q_\perp
\right)
\right|^2=\frac{d\sigma^\text{diffr}_\text{scat}}{d^2q_\perp}$, and
appropriate averaging over the atomic configurations is due:
\begin{equation}\label{av}
    \left\langle\frac{d\sigma_{\text{rad}}}{d\Gamma_{k}}\right\rangle=\left\langle\int
d^2q_\perp
\frac{d\sigma^\text{diffr}_{\text{scat}}}{d^2q_\perp}\frac{dW_{\text{dip}}}{d\Gamma_{k}}\right\rangle.
\end{equation}
For a macroscopic target, differential cross-section (\ref{av}) must
be proportional to the target area if the beam is still wider than
the target, or to the area of the beam transverse section, if it is
narrower than the target (as is a usual practice) and transversely
uniform (otherwise, we can consider any beam part uniform relative
to target inhomogeneities). Dividing (\ref{av}) by the interaction
area $S$, one obtains a quantity independent of $S$ (but
proportional to the target matter density and thickness) and having
the meaning of differential probability for the given radiative
process to occur per one particle passed through the target.

With $\frac{dW_{\text{dip}}}{d\Gamma_{k}}$ given by
Eq.~(\ref{Wdip}), in (\ref{av}) one encounters two basic integrals:
\begin{subequations}
\begin{eqnarray}
% \nonumber to remove numbering (before each equation)
  \frac1S\left\langle\int
d\sigma^\text{diffr}_{\text{scat}}\bm q_\perp^2\right\rangle\! &=&\!
\left\langle\bm q_\perp^2\right\rangle, \label{q1}\\
  \frac1S\left\langle\!\int\!\! d\sigma^\text{diffr}_{\text{scat}}(2q_{\perp
m}q_{\perp n}\!-\!\bm q_\perp^2\delta_{mn})\!\!\right\rangle\! &=&\!
\left\langle2q_{\perp m}q_{\perp n}\!-\!\bm
q_\perp^2\delta_{mn}\!\right\rangle,\nonumber\\\label{qq-qq}
\end{eqnarray}
\end{subequations}
having the meaning of average momentum squares. For our analysis to
reach beyond the conventional case of isotropic target, it is
prerequisite that average (\ref{qq-qq}) differs from zero. In
particular, that allows one to anticipate non-zero polarization of
the bremsstrahlung beam as a whole. Physically, this average is
related to the azimuthal anisotropy (``ellipticity") in scattering,
even if being not a true measure of the latter due to the radiative
character of the averaging (see Subsec.~\ref{subsec:beyondfact}).

\subsection{Vector anisotropy parameter and the anisotropy degree}
Instead of (\ref{qq-qq}), it is convenient to deal with the ratio of
(\ref{qq-qq}) to (\ref{q1}), which can serve as a direct measure of
the scattering asymmetry. This ratio, which is a symmetric traceless
tensor in 2 transverse dimensions, can be characterized by the
direction of one of its two eigenvectors and the corresponding
eigenvalue (another eigenvector will be orthogonal to the first one
and correspond to the eigenvalue opposite in sign). Let $\bm N$
stand for the eigenvector corresponding to the positive eigenvalue,
then we express
\begin{eqnarray}\label{tens=}
\frac{\left\langle2q_{\perp m}q_{\perp n}-\bm
q_\perp^2\delta_{mn}\right\rangle}{\left\langle\bm
q_\perp^2\right\rangle} \overset{\text{def}}=2N_mN_n\!-\!\bm
N^2\delta_{mn}.
\end{eqnarray}
If tensor $\left\langle q_{\perp m}q_{\perp n}\right\rangle$
diagonalizes in axes $x$, $y$, and, say, $\left\langle
q^2_y\right\rangle\geq\left\langle q^2_x\right\rangle$, then
\begin{equation}\label{N2=asymm}
    N^2=\frac{\left\langle q_y^2\right\rangle-\left\langle q_x^2\right\rangle}{\left\langle q_y^2\right\rangle+\left\langle
    q_x^2\right\rangle}
\end{equation}
(and $\bm N\parallel Oy$). That implies a constraint
\begin{equation}\label{N<1}
    \bm N^2\leq1.
\end{equation}
Covariantly, one can infer upper bound (\ref{N<1}) by squaring both
sides of (\ref{tens=}) and taking double trace.

Now, the average differential probability of  radiation can be
phrased in terms of the introduced vector $\bm N$:
\begin{eqnarray}\label{sigmarad}
  x_\omega\left\langle\frac{dW_{\text{dip}}}{dx_\omega d^2\Theta}\right\rangle=\frac{\alpha}{4\pi^2}\frac{\left\langle\bm
    q^2_\perp\right\rangle}{m^2\left(
1+\Theta^2\right)^2}\qquad\qquad\qquad\qquad\qquad\nonumber\\
\times\bigg\{2(1-x_\omega)\!\left[(G_{im}e'_i)^2(1-
N^2)\!+\!2(G_{im}N_me'_i)^2\right]\quad\quad\nonumber\\
+x_\omega^2\bigg\}.\qquad
\end{eqnarray}

At $N^2=1$, Eq.~(\ref{sigmarad}) essentially coincides with
Eq.~(\ref{Wdip}). Decomposing also the leftmost unity in braces of
Eq.~(\ref{sigmarad}) as $1\equiv(1-N^2)+N^2$, we get a
representation in form of an incoherent mixture of bremsstrahlung on
isotropic target with that on an anisotropic one, in proportion
$(1-N^2):N^2$ determiined by the target anisotropy degree. But for
polarization characteristics that superposition is non-trivial,
inasmuch as the polarization direction and degree does not express
as any simple superposition.

\subsection{Relaxing the scattering factorization assumption}\label{subsec:beyondfact}
An important concern at application of bremsstrahlung theory to
particle passage through matter is the vulnerability of the
scattering factorization condition (\ref{qzra<<1}) due to
significant target thickness. Fortunately, a way for generalization
beyond the factorization is known, which preserves the Dirac matrix
structure of the radiation matrix element, only trading the
transferred momentum $\bm q_\perp$ times $A_\text{scat}$ for some
overlap of initial and final electron wave functions, now involving
integration over longitudinal coordinates \footnote{The only
essential condition is that scattering angles do not become
comparable to $\gamma^{-1}$ due to multiple scattering within the
photon formation length (\ref{lform}).}. To make the text
self-contained, we briefly remind the idea behind that
generalization \cite{Olsen-Maximon-Wergeland}.

In the first place, it is suggestive to straightforwardly linearize
the primordial factorized matrix element (\ref{Mrad}) with respect
to $\bm q_\perp$:
\begin{eqnarray}
% \nonumber to remove numbering (before each equation)
  M_\text{rad} &=& \bar u'\bigg\{\!\left(\frac{E}{p'\cdot k}-\frac{E'}{p\cdot
  k}\right)e'^*\cdot\gamma\nonumber\\
  &\,&+\frac{e'^*\cdot\gamma q\cdot\gamma \gamma^0}{2p'\cdot k}
  +\frac{\gamma^0 q\cdot\gamma e'^*\cdot\gamma}{2p\cdot k}\bigg\}u \nonumber\\
   &\overset{\text{dip}}\approx& \frac1{q_z}\bar u'\bigg\{\frac1{q_z}\left(\bm v-\bm v'\right)\cdot \bm q_\perp \bm
   e'^*\cdot\bm\gamma\nonumber\\
   &\,&-\frac{\bm e'^*\cdot\bm\gamma \bm q_\perp\cdot\bm\gamma \gamma^0}{2E}
  -\frac{\gamma^0 \bm q_\perp\cdot\bm\gamma
  \bm e'^*\cdot\bm\gamma}{2E'}\bigg\}u.\quad\label{dipole}
\end{eqnarray}
Here we have used Eqs.~(\ref{dip-denom}) and relations \pagebreak
%\begin{subequations}
\begin{eqnarray}\label{61}
% \nonumber to remove numbering (before each equation)
  \frac{E}{p'\cdot k}-\frac{E'}{p\cdot
  k}=\frac{2EE'}{m^2\omega}\left[\frac1{1+\left(\bm\Theta+\frac{\bm q}m\right)^2}-\frac1{1+\bm\Theta^2}\right]\quad\nonumber\\
  \approx-\frac{m\omega}{EE'q_z^2}\bm\Theta\cdot\bm
  q_\perp\approx\frac1{q_z^2}\left(\bm v'-\bm v\right)\cdot
  \bm q_\perp.\quad
\end{eqnarray}
%\end{subequations}
Whichever the further method of evaluating the spin-averaged
probability, the corresponding differential probability is some
bilinear form in both $\bm e'$ and $\bm q_\perp$, and the answer is
already known -- it is of the Compton-like form (\ref{Wdip}).

To go beyond the factorization assumption, we have to start with the
exact representation of the matrix element in terms of overlap of
initial and final electron wave functions in the static field of the
target:
\begin{equation}\label{overlap-exact}
    T=\sqrt{4\pi} i e\int d^3r e^{-i\bm k\cdot\bm r}\bar\psi '(\bm
    r)\bm e'^*\cdot\bm\gamma\psi(\bm r).
\end{equation}
In the ultra-relativistic limit, the spin structure of elecron wave
functions assumes a field-independent form \cite{BLP}:
\begin{subequations}\label{FSM}
\begin{equation}\label{}
    \psi(\bm r)\approx e^{i\bm p\cdot\bm r}\left(1+\frac
    i{2E}\nabla_\perp\cdot\bm\gamma\gamma^0\right)\varphi(\bm r)u,
\end{equation}
\begin{equation}\label{}
    \bar\psi '(\bm r)\approx \bar u 'e^{-i\bm p'\cdot\bm r}\left(1-\frac
    i{2E'}\gamma^0\nabla_\perp\cdot\bm\gamma\right)\varphi'^*(\bm
    r),
\end{equation}
\end{subequations}
where modulating scalar functions $\varphi$, $\varphi'$ obey
Klein-Gordon type equations
\begin{subequations}\label{eqsforphi}
\begin{equation}\label{}
    {\bm v}\cdot\nabla\varphi(\bm r)-V(\bm r)\varphi(\bm r)=\frac1{2E}\left[\triangle-V^2(\bm r)\right]\varphi(\bm r),
\end{equation}
\begin{equation}\label{}
    -{\bm v}'\cdot\nabla\varphi'(\bm r)-\!V(\bm r)\varphi'(\bm r)=\!\frac1{2E'}\!\left[\triangle-V^2(\bm r)\right]\!\varphi'(\bm
    r)
\end{equation}
\end{subequations}
($\bm v$ and $\bm v'$ are the initial and final electron velocities
defined by (\ref{vel}), and $V(\bm r)$ the potential energy of the
electron in the electron field of the solid target).

Upon substitution of Eqs.~(\ref{FSM}) to (\ref{overlap-exact}),
\begin{eqnarray}\label{Tur}
    T_\text{u.-r.}=\sqrt{4\pi} i e\int d^3r e^{i\bm q\cdot\bm r}\bar u'\bigg\{\bm
    e'^*\!\cdot\bm\gamma \varphi'^*\varphi\qquad\qquad\qquad\nonumber\\
  +\frac{i}{2E}\varphi'^*\bm e'^*\!\cdot\bm\gamma \nabla_\perp\!\cdot\bm\gamma\gamma^0\varphi-\frac{i}{2E'}\gamma^0\nabla_\perp\varphi'^*\!\cdot\bm\gamma\bm
  e'^*\!\cdot\bm\gamma\varphi\bigg\}u,\nonumber\\
\end{eqnarray}
first term in the braces of (\ref{Tur}) appears to be
energy-suppressed, because of near orthogonality of $\bm e'^*$ to
$\bar u'\bm\gamma u$, and so gives contribution of the same order as
the second and third terms of (\ref{Tur}) (spin corrections)
containing energy denominators explicitly. The matrix element
non-factorizability implies that in spite of condition $q_z\ll
q_\perp$, one can not neglect $q_z$ component in the exponent here,
because at the scale $L_\text{corr}$ of contributing longitudinal
distances we may have $q_zL_\text{corr}\sim1$.

It seems that in Eq.~(\ref{Tur}) there are different types of
overlaps involving scalar wave functions, but in the dipole
approximation they all appear to be interrelated. Indeed,
\begin{subequations}
\begin{eqnarray}
% \nonumber to remove numbering (before each equation)
  \!\!&\,&\!\! \int d^3re^{i\bm q\cdot\bm r}\varphi'^*(\bm r)\varphi(\bm r)\nonumber\\
  \!\!&\approx&\!\! \frac i{q_z}\int d^3r e^{i\bm q\cdot\bm r}({\bm
v}'\cdot\nabla)\left\{\varphi'^*(\bm r)\varphi(\bm r)\right\}\nonumber\\
  \!\!&\equiv&\!\! \frac i{q_z}\int d^3r e^{i\bm q\cdot\bm
r}\varphi'^*(\bm r)\left({\bm v}'-{\bm
v}\right)\cdot\nabla\varphi(\bm r)\nonumber\\
  \!\!&\,&\!\!+\frac i{q_z}\int d^3r e^{i\bm q\cdot\bm r}\left[({\bm
v}'\cdot\nabla\varphi'^*)\varphi+\varphi'^*{\bm
v}\cdot\nabla\varphi\right]\nonumber\\
  \!\!&=&\!\!\frac i{q_z}\!\int d^3re^{i\bm q\cdot\bm
r}\!\left\{\varphi'^*\left(\bm v'-\bm
v\right)\!\cdot\!\nabla\varphi+\mathcal{O}\left(\frac{q^2+V^2}{E'}\varphi'^*\varphi\right)\!\right\}\nonumber\\ \label{66a}\\
  \!\!&=&\!\!\frac i{q_z}\left(\bm v'-\bm v\right)\!\cdot\!\!\int d^3re^{i\bm
q\cdot\bm r} \varphi'^*\nabla_\perp\varphi
\left\{1+\mathcal{O}\!\left(\frac{q_\perp}{x_\omega
m}\right)\!\right\}\!,\quad\label{phiphitransform}
\end{eqnarray}
\end{subequations}
where in passing to Eq.~(\ref{66a}) we have used wave equations
(\ref{eqsforphi}), and in passing to Eq.~(\ref{phiphitransform}) --
the estimate $\left|\bm v'-\bm v\right|\sim\frac{x_\omega m}{E'}$
(cf. Eq.~(\ref{61})). Another type of overlap
\begin{eqnarray}\label{}
% \nonumber to remove numbering (before each equation)
  &\,&\int d^3re^{i\bm q\cdot\bm r}\left(\nabla_\perp\varphi'^*(\bm r)\right)\varphi(\bm r)\nonumber\\
  &=& -\int d^3re^{i\bm q\cdot\bm r}\varphi'^*\nabla_\perp\varphi-i\bm q_\perp\int d^3r e^{i\bm q\cdot\bm r}\varphi'^*\varphi\nonumber\\
  &\approx&\left[-1+\frac{\bm q_\perp}{q_z}\left(\bm v'-\bm v\right)\cdot\right]\int d^3re^{i\bm q\cdot\bm r}\varphi'^*\nabla_\perp\varphi \nonumber\\
  &=&-\int d^3re^{i\bm q\cdot\bm r}\varphi'^*\nabla_\perp\varphi\left\{1+\mathcal{O}\left(\frac{q_\perp}{m}\right)\right\}.
\end{eqnarray}
Thus, the entire overlap (\ref{Tur}) can be cast in terms of a
single overlap between the scalar wave functions
\begin{equation}\label{Idef}
    \mathfrak{\bm I}_\perp\left(q_z,\bm q_\perp\right)=q_z\int d^3re^{iq_zz+i\bm q_\perp\cdot\bm
r_\perp}\varphi'^*(\bm r)\nabla_\perp\varphi(\bm r),
\end{equation}
specifically:
\begin{eqnarray}\label{Tdip}
% \nonumber to remove numbering (before each equation)
  T_\text{dip} = \sqrt{4\pi}e\frac1{q_z}\bar u'\bigg\{\frac1{q_z}\left(\bm v-\bm v'\right)\cdot\mathfrak{\bm I}_\perp\bm e'^*\cdot\bm\gamma\qquad\qquad\nonumber\\
  -\frac1{2E}\bm e'^*\cdot\bm\gamma \mathfrak{\bm I}_\perp\cdot\bm\gamma\gamma^0-\frac1{2E'}\gamma^0\mathfrak{\bm I}_\perp\cdot\bm\gamma\bm
  e'^*\cdot\bm\gamma\bigg\}u.
\end{eqnarray}
This is observed to have the very same Dirac matrix structure as
(\ref{dipole}), only with $\mathfrak{\bm I}_\perp$ emerging in place
of $\bm q_\perp$.

As we see, the recipe for the generalization beyond the scattering
factorization is to make in the factorized matrix element
\begin{equation*}
    T_\text{fact}=\sqrt{4\pi}eA^\text{diffr}_\text{scat}(\bm q_\perp)M_\text{rad},
\end{equation*}
with $M_\text{rad}$ given by (\ref{dipole}), a replacement
\begin{equation}\label{fact subst}
    \bm q_\perp A^\text{diffr}_{\text{scat}}\left(\bm q_\perp\right)\overset{q_zL_{\text{corr}}\sim1}
    \to \mathfrak{\bm I}_\perp\left(q_z,\bm q_\perp\right),\qquad
q_z=q_z\left(\omega,\bm\Theta\right).
\end{equation}
Here, factor $e^{iq_zz}$ represents the effects of longitudinal
coherence sensitivity. Correspondence with the scattering
factorization is achieved when this exponential can be put to unity
(after a preliminary integration over $z$ by parts, to make the
integrand vanish at infinity):
\begin{eqnarray}
% \nonumber to remove numbering (before each equation)
  \mathfrak{\bm I}_\perp = i\int d^2r_\perp e^{i\bm
q_\perp\cdot \bm r_\perp}\int dz e^{iq_zz}\frac\partial{\partial
z}\left[\varphi'^*(\bm r)\left(\nabla_\perp\right)\varphi(\bm
r)\right]\nonumber\\
% \nonumber to remove numbering (before each equation)
\overset{q_zL_{\text{corr}}\ll1}\to i\int d^2r_\perp e^{i\bm
q_\perp\cdot \bm r_\perp} 1(\nabla_\perp)\varphi(\bm r_\perp,z\gg
L_\text{corr})\qquad\nonumber\\
=\bm q_\perp\int d^2r_\perp e^{i\bm q_\perp\cdot \bm
r_\perp}\varphi(\bm r_\perp,z\gg
L_\text{corr}).\qquad\qquad\quad\label{Ilim}
\end{eqnarray}
By the Huygens principle (see, e. g., \cite{Huygens}), the latter
integral equals to the elastic scattering amplitude, if normalized
as in Eq.~(\ref{Anorm}). So,
\begin{equation}\label{fact change}
    \mathfrak{\bm I}_\perp\overset{q_zL_{\text{corr}}\ll1}\to\bm q_\perp A^\text{diffr}_{\text{scat}}(\bm
    q_\perp),
\end{equation}
offering a check of the replacement rule consistency (\ref{fact
subst}).

To obtain the spin-averaged probability corresponding to the
generalized matrix element (\ref{Tdip}), one needs no special
calculation. Obviously, it is a bilinear form in $\mathfrak{\bm
I}_\perp$, which can be retrieved from the factorized bilinear form
by replacement $d\sigma^\text{diffr}_\text{scat}q_{\perp m}q_{\perp
n}\to\frac{d^2q_\perp}{(2\pi)^2}\mathfrak I_{\perp m}\mathfrak
I_{\perp n}$. Thereat, the basic averages for our quadratic form
promote from (\ref{q1}, \ref{qq-qq}) to
\begin{subequations}
\begin{equation}\label{q^2 rad}
\frac1S\left\langle\int d\sigma^\text{diffr}_{\text{scat}}\bm
q_\perp^2\right\rangle\overset{q_zL_{\text{corr}}\sim1}\to\frac1S\left\langle\int\frac{d^2q_\perp}{(2\pi)^2}\left|\mathfrak{\bm
I}_\perp\right|^2\right\rangle,
\end{equation}
\begin{eqnarray}\label{tens rad}
\frac1S\left\langle\int d\sigma^\text{diffr}_{\text{scat}}(2q_{\perp
m}q_{\perp n}-\bm q_\perp^2\delta_{mn})\right\rangle\qquad\qquad\quad\nonumber\\
\overset{q_zL_{\text{corr}}\sim1}\to\frac1S\left\langle\int\frac{d^2q_\perp}{(2\pi)^2}\left(2\mathfrak
I_{\perp m}\mathfrak I_{\perp n}-\delta_{mn}\mathfrak{\bm
I}_\perp^2\right)\right\rangle.
\end{eqnarray}
\end{subequations}
The factor $1/S$ in the right-hand sides may be explicitly canceled
if $\varphi$ (but not $\varphi'$) is substituted by a normalized
wave packet in transverse coordinates.

For the modified averages (\ref{q^2 rad}-\ref{tens rad}) we
introduce same shorthands as (\ref{q1}-\ref{qq-qq}) but with the
subscript ``rad":
\begin{eqnarray*}
% \nonumber to remove numbering (before each equation)
  \frac1S\left\langle\int\frac{d^2q_\perp}{(2\pi)^2}\left|\mathfrak{\bm
I}_\perp\right|^2\right\rangle\equiv\left\langle\bm
q_\perp^2\right\rangle_{\text{rad}}, \\
  \frac1S\left\langle\int\frac{d^2q_\perp}{(2\pi)^2}\left(2\mathfrak
I_{\perp m}\mathfrak I_{\perp n}-\delta_{mn}\mathfrak{\bm
I}_\perp^2\right)\right\rangle\nonumber\\
\equiv\left\langle2q_{\perp m}q_{\perp n}-\bm
q_\perp^2\delta_{mn}\right\rangle_{\text{rad}}.
\end{eqnarray*}
For expression of the anisotropy parameter, equations (\ref{tens=},
\ref{N2=asymm}) remain valid, only with the replacement
$\left\langle\bm q_\perp^2\right\rangle\to\left\langle\bm
q_\perp^2\right\rangle_{\text{rad}}$:
\begin{eqnarray}
\frac{\left\langle2q_{\perp m}q_{\perp n}-\bm
q_\perp^2\delta_{mn}\right\rangle_{\text{rad}}}{\left\langle\bm
q_\perp^2\right\rangle_{\text{rad}}}=2N_mN_n\!-\!\bm
N^2\delta_{mn},\label{tens=2}\\
N^2=\frac{\left\langle q_y^2\right\rangle_\text{rad}-\left\langle
q_x^2\right\rangle_\text{rad}}{\left\langle
q_y^2\right\rangle_\text{rad}+\left\langle
    q_x^2\right\rangle_\text{rad}}.\qquad\qquad\label{N2=asymm2}
\end{eqnarray}

\section{The case of azimuthally anisotropic scattering at electron passage through an oriented crystal}\label{subsubsec:quant}
Among conceivable applications of the connection between the target
intrinsic anisotropy and radiation polarization is the possibility
of preparation of a polarized photon beam. For reliability of
polarization asymmetry measurements, the beam polarization degree
must be high enough, at least a few tens percent, and thence about
as high must be $N^2$. It is not, however, obvious, whether that
sizeable $N^2$ can be attained with \emph{macroscopic} targets. As
we had mentioned in the Introduction, the main obstacle thereto is
the hard isotropic contribution in scattering. When treating
interaction with individual atom as perturbative, in (\ref{av}) in
the integral over $\bm q_\perp$, or, in an equivalent integral over
impact parameters, the contribution from the atomic distance scale
$\sim r_\text{a}$ is comparable to that from the distances from the
nucleus of the order $\sim m^{-1}$, where the impact area is smaller
but the acting force, and the generated radiation, is stronger. That
familiarly leads to a logarithmic divergence of the integral over
$d^2q_\perp$ from
$\frac{d\sigma_\text{rad}}{d^2q_\perp}=\frac{d\sigma_\text{scat}}{d^2q_\perp}dW_\text{rad}$,
with $dW_\text{rad}\propto \bm q_\perp^2$ and
$\frac{d\sigma_\text{scat}}{d^2q_\perp}\underset{q_\perp\gg
r_\text{a}^{-1}}\sim\bm q_\perp^{-4}$ (Rutherford tail). Introducing
appropriate cutoffs, upper one due to the dipole approximation
failure at $q_\perp\sim m$, and lower one due to the atomic
form-factor regulation, one gets with the logarithmic accuracy
\begin{equation}\label{LogAppr}
    \int d^2q_\perp \bm q_\perp^2\frac1{\bm
    q_\perp^4}=\pi\int_{q_{\perp\text{min}}=r_\text{a}^{-1}}^{q_{\perp\text{max}}=m}\frac{dq_\perp}{q_\perp}=\pi\ln
    mr_\text{a}\approx\pi\ln\frac1\alpha.
\end{equation}
Since in vicinities of the nuclei the scattering is isotropic, the
anisotropy parameter $N^2$ gets suppressed at least by a factor of
$\ln\frac1\alpha\approx 5$.

\begin{figure}
\includegraphics{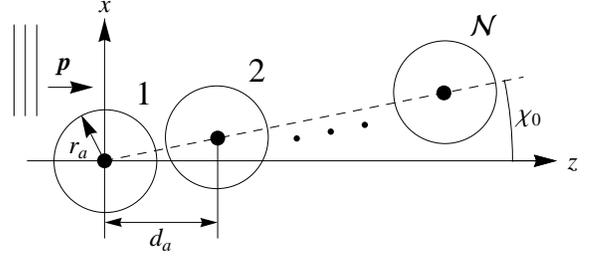}
\caption{\label{fig:atomicstring} Electron small-angle passage
through an atomic row. Momentum transfers in $y$ direction
(transverse to the picture) are enhanced compared to those in
$x$-direction.}
\end{figure}

A remedy to the encountered suppression could be sought in utilizing
oriented crystals. Once one aligns some strong crystallographic axis
by a small angle $\chi_0\ll 1$ relative to the electron incidence
direction (see Fig.~\ref{fig:atomicstring}), the isotropy may
persist only up to the distance of \emph{transverse} separation of
atomic nuclei in the string, $\Delta r_\perp\sim d_\text{a}\chi_0$,
where $d_\text{a}>2r_\text{a}$ is the distance between atomic nuclei
in the row, while from scale $\Delta r_\perp$ up to $r_\text{a}$ the
scattering should become anisotropic (stronger in the direction
transverse to the beam-string plane). At a scale greater than
$\chi_0d_\text{a}$, the cross-section will no longer be a sum of
logarithmic cross-sections of scattering on individual atoms, but,
rather, the motion will be governed by the aggregate potential of
the atoms. The number of atoms overlapping at a given impact
parameter is $\sim\frac{r_\text{a}}{\chi_0d_\text{a}}$, and this is
the factor the cross-section must increase by, whilst Coulombic
logarithms do not develop anymore in this region.

\subsection{Electron interaction with a single atomic string}

To verify the above assumption, yet to get an idea of the
longitudinal coherence sensitivity, consider first a problem of
electron radiation at scattering on a single atomic row under a
small angle of incidence. In capacity of initial and final state
wave functions in the atomic row potential $V(\bm r)$, take for
simplicity the eikonal approximation (corresponding to the neglect
in (\ref{eqsforphi}) of the right-hand sides, as well as neglect of
the angle between $\bm v$ and $\bm v'$):
\begin{subequations}\label{eik}
\begin{equation}\label{}
    \varphi(\bm r) \approx e^{-i\int^z_{-\infty}dz'V(z',\bm r_\perp)},
\end{equation}
\begin{equation}\label{}
    \varphi'^*(\bm r) \approx e^{-i\int^\infty_z dz'V(z',\bm r_\perp)},
\end{equation}
\end{subequations}
\begin{equation}\label{}
    \varphi'^*(\bm r)\varphi(\bm r)=e^{i\chi_0(\bm r_\perp)},\quad\chi_0(\bm
  r_\perp)=-\int^\infty_{-\infty}dzV\left(z',\bm r_\perp\right).
\end{equation}
Here $\chi_0(\bm r_\perp)$ is commonly called the eikonal phase.
Substituting (\ref{eik}) into (\ref{Idef}), one obtains
\cite{BKStrakh}
\begin{eqnarray}
% \nonumber to remove numbering (before each equation)
  \mathfrak{\bm I}_\perp  = -iq_z\int d^3re^{i\bm q\cdot\bm r+i\chi_0(\bm r_\perp)}\nabla_\perp\int^z_{-\infty}dz'V(z',\bm r_\perp)\,\,\nonumber\\
   = \int d^2r_\perp e^{i\bm q_\perp\cdot\bm r_\perp+i\chi_0(\bm r_\perp)}\nabla_\perp\int^\infty_{-\infty}dze^{iq_zz}V(z,\bm
   r_\perp)\quad
\end{eqnarray}
(the second equality results after integration by parts over $z$).
The integrals of $|\mathfrak{\bm I}_x(\bm q_\perp)|^2$,
$|\mathfrak{\bm I}_y(\bm q_\perp)|^2$ engaged in definition of $N^2$
evaluate in a particularly simple form:
\begin{eqnarray}\label{}
    \int d^2q_\perp|\mathfrak{\bm I}_{x,y}|^2&=&(2\pi)^2\int
    d^2r_\perp\left|\nabla_{x,y}\int^\infty_{-\infty}dze^{iq_zz}V\right|^2\nonumber\\
    &=&\int d^2q_\perp\bm q^2_{x,y}\left|\int d^3re^{i\bm q\cdot\bm r}V(\bm
    r)\right|^2.
\end{eqnarray}
I. e., the eikonal phase actually does not contribute to the given
integral, and the result is equivalent to the Born approximation.

To proceed, we have to specify the atomic row potential. Let us, for
simplicity, model it by a superposition of individual screened
Coulombic potentials. For ia row of $\mathcal{N}$ identical atoms
with nucleus charge $Z$, located in the $xz$ plane at an angle
$\chi_0\ll1$ to $z$-axis,
\begin{equation}\label{ReduceToBorn}
    V_\text{row}(\bm
    r)=Z\alpha\sum^{\mathcal{N}-1}_{n=0}\frac{e^{-\frac{y^2+(x-\chi_0d_\text{a}n)^2+(z-d_\text{a}n)^2}{r_\text{a}}}}{\sqrt{y^2+(x-\chi_0d_\text{a}n)^2+(z-d_\text{a}n)^2}}.
\end{equation}
Fourier transform thereof results as
\begin{eqnarray}\label{}
    \int d^3re^{i\bm q\cdot\bm r}V_\text{row}(\bm r)=\frac{4\pi Z\alpha}{\bm
    q_\perp^2+r_\text{a}^{-2}}\sum^{\mathcal{N}-1}_{n=0}e^{i(q_z+\chi_0q_x)d_\text{a}n}\qquad\nonumber\\
    =\frac{4\pi Z\alpha}{\bm
    q_\perp^2+r_\text{a}^{-2}}e^{i(q_z+\chi_0q_x)d_\text{a}(\mathcal{N}-1)/2}\frac{\sin\frac{(q_z+\chi_0q_x)d_\text{a}\mathcal{N}}2}{\sin\frac{(q_z+\chi_0q_x)d_\text{a}}2},\nonumber
\end{eqnarray}
where in the first denominator term $q_z^2$ had been neglected
compared to $r_\text{a}^{-2}$. When squaring (\ref{ReduceToBorn}),
the sine ratio factor in a familiar way may be approximated by a
sequence of equidistant $\delta$-functions:
\begin{eqnarray}\label{}
    \frac{\sin^2\frac{(q_z+\chi_0q_x)d_\text{a}\mathcal{N}}2}{\sin^2\frac{(q_z+\chi_0q_x)d_\text{a}}2}\underset{\mathcal{N}\gg
    1}{\approx}\pi \mathcal{N}\sum^\infty_{j=-\infty}\delta\left(\frac{(q_z+\chi_0q_x)d_\text{a}}2-\pi
    j\right)\nonumber\\
    =\frac{2\pi \mathcal{N}}{\chi_0d_\text{a}}\sum^\infty_{j=-\infty}\delta\left(q_x-\frac{2\pi
    j}{\chi_0d_\text{a}}+\frac{q_z}{\chi_0}\right),\qquad\nonumber
\end{eqnarray}
i. e., $q_x$-integration reduces to summation over 1-dimensional
inverse lattice vectors. As for $q_y$-integration, it involves 2
basic integrals:
\begin{subequations}
\begin{equation}\label{}
    \int^\infty_{-\infty}\frac{dq_y}{\left(q_y^2+q_x^2+r_\text{a}^{-2}\right)^2}=\frac{\pi}{2\left(q_x^2+r_\text{a}^{-2}\right)^{3/2}}
\end{equation}
for $\int d^2q_\perp \left|\mathfrak{\bm I}_x\right|^2$, and
\begin{equation}\label{}
    \int^\infty_{-\infty}\frac{dq_yq_y^2}{\left(q_y^2+q_x^2+r_\text{a}^{-2}\right)^2}=\frac{\pi}{2\sqrt{q_x^2+r_\text{a}^{-2}}}
\end{equation}
\end{subequations}
for $\int d^2q_\perp \left|\mathfrak{\bm I}_y\right|^2$. In the
final result, it is convenient to treat in the sum the term $j=0$
separately. It has the meaning of contribution from `continuous'
potential, constant along the string. In the higher terms one may
neglect $r_\text{a}^{-2}$ relative to
$\left(\frac{2\pi}{\chi_0d_\text{a}}\right)^2$. Thereby one obtains
\begin{subequations}
\begin{equation}\label{intJ2y}
    \frac{1}{\pi \mathcal{N}\left(4\pi Z\alpha\right)^2}\!\int\! d^2q_\perp \left|\mathfrak{\bm
    I}_y\right|^2
    =\frac{\pi}{\chi_0d_\text{a}\sqrt{\frac{q_z^2}{\chi_0^2}+r_\text{a}^{-2}}}
    +\sum_{j=1}^{j_{\max}}\frac1j,
\end{equation}
\begin{equation}\label{intJ2x}
    \frac{1}{\pi \mathcal{N}\left(4\pi Z\alpha\right)^2}\!\int\! d^2q_\perp \left|\mathfrak{\bm
    I}_x\right|^2
    =\frac{\pi q_z^2}{\chi_0^3d_\text{a}\left(\frac{q_z^2}{\chi_0^2}+r_\text{a}^{-2}\right)^{\!3/2}}
    +\sum_{j=1}^{j_{\max}}\frac1j.
\end{equation}
\end{subequations}
The summation upper limit $j_{\max}$ is determined by the same
principle as that of integration in (\ref{LogAppr}) -- it is set at
$q_x\sim m$, so
\begin{equation*}\label{}
    j_{\max}=j_{\max}(\chi_0)\sim\frac{\chi_0d_\text{a}}{2\pi}m\sim 20\,\chi_0,
\end{equation*}
\begin{equation*}\label{}
    \sum_{j=1}^{j_{\max}(\chi_0)}\frac1j\approx\max\left\{1,\ln\left[
    e^{C_E}j_{\max}(\chi_0)\right]\right\}\overset{\text{def}}{=}L_0\,(\chi_0).
\end{equation*}

For scattering on a single string at zero temperature, no additional
averaging over string ensembles is required, and substitution of
(\ref{intJ2y}, \ref{intJ2x}) to (\ref{N2=asymm2}) gives the result
for the azimuthal anisotropy parameter
\begin{equation}\label{rowN2}
    N^2=\frac{1}{1+2\frac{q_z^2r_\text{a}^2}{\chi_0^2}+\frac2\pi
    \left(1+\frac{q_z^2r_\text{a}^2}{\chi_0^2}\right)^{3/2}\frac{\chi_0d_\text{a}}{r_\text{a}}L_0(\chi_0)}
\end{equation}
(we remind that $q_z\left(\omega,\bm\Theta\right)$ is defined by
Eq.~(\ref{qz})).

Based on the above explicit formula, let us now assess the effects
of scattering on individual nuclei and of the longitudinal coherence
sensitivity. In the denominator of (\ref{rowN2}), terms
$\frac{q_z^2r_\text{a}^2}{\chi_0^2}$ reflect the effect of coherence
sensitivity on the anisotropy of radiation distribution. Apparently,
increase of $q_z$ through $\omega$ always suppresses the anisotropy.
The last term, containing $L_0$, accounts for effects of the string
discreteness, which are also suppressing the anisotropy. But due to
the factor $\frac{\chi_0d_\text{a}}{r_\text{a}}$ (inverse coherence
enhancement factor), this effect weakens as $\chi_0$ decreases, and
even at angles as large as
\begin{equation}\label{chi02}
\chi_0\sim0.2\,\text{rad}\sim10^\circ,\qquad L_0(\chi_0)\simeq2
\end{equation}
one has $N^2\sim0.5$, provided
$\frac{q_zr_\text{a}}{\chi_0}<\frac12$. In fact, at impact angles
(\ref{chi02}), and $q_z\sim x_\omega m/\gamma$, the ratio
$\frac{q_zr_\text{a}}{\chi_0}\sim\frac{x_\omega}{2\alpha\gamma\chi_0}$
quantifying the longitudinal coherence effect on the anisotropy,
will be small if
\begin{equation}\label{Nconst}
    \gamma\geq\frac{x_\omega}{\alpha \chi_0}10^3x_\omega \quad \Rightarrow \quad N\approx
    N(q_z\to0)=\mathrm{const}.
\end{equation}
In this case, $N^2$ can be regarded as independent of $q_z$, and
therethrough of the emitted photon momentum.

From the estimate (\ref{chi02}) one can further infer the sufficient
crystal quality, the crystal orientation precision and the beam
collimation degree; we will not discuss these items in detail
herein.

\subsection{Multiple scattering on atomic strings}
To be realistic, at electron passage through a real oriented
crystal, interaction with one string is not the whole story but only
an elementary act. Multiple interactions can affect the distribution
function in scattering angles, and yet, in case of periodical string
hitting, modify the radiation spectrum through the periodic
structure form-factor.

For successive scattering on atomic rows with nearly continuous
string potentials, the modulus of the angle between the row and the
electron motion direction is actually approximately conserved
(transverse energy conservation). Hence, in multiple scattering on
mutually parallel strings the electron momentum will diffuse over a
cone with the axis along the string direction (`doughnut' scattering
\cite{doughnut}), and for a sufficiently thick target the scattering
must isotropize. To keep the scattering azimuthal anisotropy
significant, one should not permit the electron passage to such late
a stage. The allowable target thickness $L$ is estimated by assuming
that at this distance the electron interacts with
$L\frac{\chi_0}{d_\text{a}}$ strings, scattering on each one through
a small angle
\begin{equation*}\label{}
    \chi_1\sim\frac{F_0}E\frac{r_\text{a}}{\chi_0}=\frac{2V_0}{E\chi_0},
\end{equation*}
whence the change of the azimuth $\frac{\chi_1}{\chi_0}$ needs be
$\ll1$, too. Then, the mean square of the scattering azimuthal angle
on a sequence of (statistically independent) strings,
$L\frac{\chi_0}{d_\text{a}}\frac{\chi_1^2}{\chi_0^2}$, is required
to be less than unity. So, the condition for the target thickness
ensues as
\begin{equation}\label{}
    L<L_\text{isotr}\!\sim d_\text{a}\frac{\chi_0}{\chi_1^2}\sim d_\text{a}\frac{E^2\chi_0^3}{4V_0^2}\sim
    \frac{d_\text{a}}{4\alpha^4}\chi_0^3\gamma^2\quad\left(\frac{m}{V_0}\!\sim\!\frac1{\alpha^2}\right),
\end{equation}
where $\frac{d_\text{a}}{4\alpha^4}\sim \mathrm{1-2\,cm}$. At
$\gamma$ satisfying (\ref{Nconst}), with $\chi_0$ of the order of
(\ref{chi02}), the effect of doughnut isotropization is weak.

On the other hand, if strings are encountered along the particle
path periodically (`string of strings' radiation \cite{doughnut},
similar to coherent bremsstrahlung \cite{Diambrini}), then even at
$\frac{q_zr_\text{a}}{\chi_0}<\frac12$ one can still have
$\frac{q_zd_\text{s}}{\chi_0}\sim1$, with $d_\text{s}$ the distance
between the strings. Then, coherence effects in radiation may
develop on a larger spatial scale. If the period of string sequence
is equal to the photon formation length (\ref{lform}) at some
$\omega$, $\theta$, then the spectrum contains a resonance radiation
peak at this frequency. Within the peak, the value of $q_z$, and
therewith of $\bm q_\perp$, may be regarded as certain (the `point
effect' in coherent bremsstrahlung \cite{Diambrini}). Then, the
azimuthal asymmetry degree, again, would approach unity, minus
corrections on thermal atom oscillations, lattice defects, etc.

To conclude this section, let us remark that physically interesting
examples of media with non-zero $N^2$, of course, are not restricted
to string-like configurations described above. There are many other
configurations (textured polycrystals, bent crystals
\cite{BondCBBC}, polarized non-spherical nuclei, etc.), which even
if aren't particularly convenient for polarized photon beam
production, but are physically interesting by themselves. Such
systems allow diagnostics by the (polarized) bremsstrahlung. The
bremsstrahlung properties for such systems must be qualitatively
similar, since they are characterized by the aggregate vector $\bm
N$ only. Analysis of these properties depending on the parameter
$N$, as it assumes intermediate values $0<N<1$, is carried out in
the next section.

\section{Bremsstrahlung polarization observables}\label{sec:ang}

In general, function $\bm N\left(\omega,\Theta\right)$ is
model-dependent, and models, in principle, may vary. To derive
model-independent conclusions, let us for the rest of this paper
assume $\bm N$ to be a constant (see condition (\ref{Nconst})),
though arbitrary parameter. Therewith, we will investigate influence
of $\bm N$ on the radiation intensity and on polarization, both in
the integral photon beam and in the detail of angular distribution.

Upon replacement in (\ref{sigmarad}) $\left\langle\bm
    q^2_\perp\right\rangle\to\left\langle\bm
    q^2_\perp\right\rangle_{\text{rad}}$, the radiation intensity
differential distribution reads
\begin{eqnarray}\label{sigmarad2}
  x_\omega\left\langle \frac{dW_{\text{dip}}}{dx_\omega d^2\Theta}\right\rangle=\frac{\alpha}{4\pi^2}\frac{\left\langle\bm
    q^2_\perp\right\rangle_{\text{rad}}}{m^2\left(
1+\Theta^2\right)^2} \qquad\qquad\qquad\quad\nonumber\\
\times\!\left\{{2(1\!-\!x_\omega)}\!\left[(G_{im}e'_i)^2(1\!-\!
N^2)\!+\!2(G_{im}N_me'_i)^2\right]\!+\!x_\omega^2\right\}\!.\nonumber\\
\end{eqnarray}
To separate the unpolarized part and the polarization, one needs to
split the dependence of Eq.~(\ref{sigmarad}) on $\bm e'$ into the
isotropic and the quadrupole parts, writing
\begin{eqnarray*}
(G_{im}e'_i)^2&=&1-\frac{4(\bm\Theta\cdot\bm e')^2}{\left( 1+
\Theta^2\right)^2}\\
&\equiv&\frac{1+\Theta^4+2\left(\Theta^2\delta_{ij}-2\Theta_i\Theta_j
\right)e'_ie'_j}{(1+\Theta^2)^2},
\end{eqnarray*}
and
\begin{eqnarray*}
2(G_{im}N_me'_i)^2\qquad\quad\qquad\quad\qquad\quad\qquad\quad\qquad\quad\qquad\quad\nonumber\\
\equiv(G_{lm}N_m)^2
+\left[2G_{im}N_mG_{jn}N_n-(G_{lm}N_m)^2\delta_{ij} \right]e'_ie'_j,\nonumber\\
\end{eqnarray*}
where
\[
(G_{lm}N_m)^2=N^2-\frac{4(\bm
N\cdot\bm\Theta)^2}{\left(1+\Theta^2\right)^2}.
\]
As a result, we bring Eq.~(\ref{sigmarad2}) to the form
\begin{eqnarray}\label{2Tij-Tdeltaij}
x_\omega\left\langle \frac{dW_{\text{dip}}}{dx_\omega
d^2\Theta}\right\rangle
=\frac12x_\omega\left\langle \frac{dW_{\text{unpol}}}{dx_\omega d^2\Theta}\right\rangle\qquad\qquad\qquad\qquad\nonumber\\
+\frac{\alpha}{4\pi^2}\frac{\left\langle\bm
    q^2_\perp\right\rangle_{\text{rad}}}{m^2\left(
1+\Theta^2\right)^2}\frac{2(1-x_\omega)}{x_\omega^2}\left(2T_{ij}-T\delta_{ij}
\right)e'_ie'_j,\quad
\end{eqnarray}
with tensor $T_{ij}$ emerging as
\begin{equation}\label{Tij def}
T_{ij}= -\frac{2(1-N^2)}{\left( 1+
\Theta^2\right)^2}\Theta_i\Theta_j+G_{im}N_mG_{jn}N_n ,
\end{equation}
$T$ its trace
\begin{equation}
T=T_{ii}=-\frac{2(1-N^2)\Theta^2}{\left( 1+
\Theta^2\right)^2}+(G_{lm}N_m)^2 ,
\end{equation}
and the unpolarized part of Eq.~(\ref{2Tij-Tdeltaij}) (equal to
Eq.~(\ref{sigmarad2}) summed up over the independent polarization
states $e'_n$)
\begin{eqnarray}\label{sigmaunpol}
x_\omega\left\langle \frac{dW_{\text{unpol}}}{dx_\omega
d^2\Theta}\right\rangle=x_\omega\sum_{\bm
e'}\left\langle \frac{dW_{\text{dip}}}{dx_\omega d^2\Theta}\right\rangle\quad\qquad\qquad\qquad\nonumber\\
=\frac\alpha{2\pi^2}\frac{\left\langle\bm
    q^2_\perp\right\rangle_{\text{rad}}}{m^2\left(
1+\Theta^2\right)^2}\qquad\qquad\qquad\qquad\qquad\qquad\qquad\nonumber\\
\times\!\left\{2(1-x_\omega)\frac{1+\Theta^4+2\left[N^2\Theta^2-2(\bm
N\cdot\bm\Theta)^2 \right]}{\left(
1+\Theta^2\right)^2}+x_\omega^2\!\right\}\!.\nonumber\\
\end{eqnarray}

\subsection{Spectrum and net polarization of the integral radiation cone}\label{sec:aggr}
If angular resolution of the emitted
radiation is not pursued in the experiment (which may become
impractical at $\gamma>10^{4}$), and only the natural collimation
due to emission from an ultra-relativistic particle is utilized, one
must integrate Eq.~(\ref{2Tij-Tdeltaij}) over the small radiation
angles, i. e., a $\bm\Theta$ plane. The angular integrations are
carried out with the aid of the basic integrals
\[
\int \frac{d\phi_{\bm{\Theta}}}{2\pi}\Theta_i
\Theta_j=\frac{1}{2}\delta_{ij}\Theta^2,
\]
\begin{equation}
\int \frac{d\phi_{\bm\Theta}}{2\pi}\Theta_i \Theta_j\Theta_l
\Theta_m=\frac18\left(\delta_{ij}\delta_{lm}+\delta_{il}\delta_{jm}+\delta_{im}\delta_{jl}
\right)\Theta^4,
\end{equation}
\begin{equation}\label{theta^2int}
\int_0^\infty\frac{d\Theta^2(\Theta^2)^m}{\left(1+\Theta^2\right)^{2+n}}=\frac{m!(n-m)!}{(n+1)!}.
\end{equation}
The result is %\footnote{The salient factor 1/3 emerges from (\ref{theta^2int}) at $n=2$.}
\begin{eqnarray}\label{sigmaint}
x_\omega\left\langle\frac{dW_{\text{dip}}}{dx_\omega}\right\rangle=\frac12\int
d\Theta^2\frac{d\phi_{\bm\Theta}}{2\pi}x_\omega\left\langle\frac{dW_{\text{dip}}}{dx_\omega d^2\Theta}\right\rangle\qquad\quad\quad\nonumber\\
=\frac{\alpha\left\langle\bm
q_\perp^2\right\rangle_{\text{rad}}}{4\pi
m^2}\!\left\{\frac23(1-x_{\omega})\!\left[2+2(\bm N\cdot\bm
e')^2\!-\!N^2\right]+x_\omega^2\right\}\!.\nonumber\\
\end{eqnarray}

\begin{figure}
\includegraphics{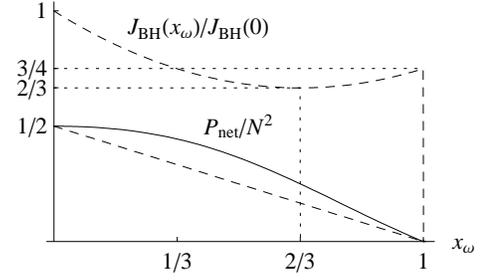}
\caption{\label{fig:Net} Dashed curve -- Bethe-Heitler spectral
distribution of bremsstrahlung energy (Eq.~(\ref{BH-our})),
normalized to its value at $x_\omega=0$. Oblique dashed line -- the
polarized fraction of the radiation spectrum divided by $N^2$, for
the case of constant $\bm N$. Solid curve -- net polarization of the
photon beam (Eq.~(\ref{Paggr})) divided by $N^2$. (Polarization
orientation is $\parallel\bm N$). }
\end{figure}

The unpolarized, Bethe-Heitler's spectral intensity (which otherwise
might be obtained by integrating Eq.~(\ref{sigmaunpol})) ensues
\begin{subequations}
\begin{eqnarray}\label{BH-our}
J_{\mathrm{BH}}\left(x_\omega\right)&=&x_\omega\left\langle\frac{dW_{\text{unpol}}}{dx_\omega}\right\rangle
=x_\omega\sum_{\bm{e}'}\left\langle\frac{dW_{\text{dip}}}{dx_\omega}\right\rangle\nonumber\\
&=&\frac{\alpha\left\langle\bm
q_\perp^2\right\rangle_{\text{rad}}}{2\pi
m^2}\left\{\frac43\left(1-x_\omega\right)+x_\omega^2
\right\}\qquad\quad\\
&\equiv&\frac{\alpha\left\langle\bm
q_\perp^2\right\rangle_{\text{rad}}}{2\pi
m^2}\frac{E'}{E}\left(\frac{E'}{E}+\frac{E}{E'}-\frac{2}{3}\right)\!\label{BH}
\end{eqnarray}
\end{subequations}
(see Fig.~\ref{fig:Net}, upper dashed curve). Therein the dependence
on $\bm N$ completely drops out -- quite naturally, recalling that
$\bm N$ is representative of the quadrupole dependence on $\bm
q_\perp$, while after integration over $\bm\Theta$ it can only be
contracted with the quadrupole tensor dependence on $\bm e'$ (as
Eq.~(\ref{sigmaint}) indicates), but after summation over $\bm e'$
all that averages to zero. The integral of the $x_\omega$-dependent
expression in Eq.~(\ref{BH-our}) is unity:
\begin{equation*}\label{}
    \int_0^1dx_\omega\left\{\frac43\left(1-x_\omega\right)+x_\omega^2\right\}=1.
\end{equation*}

Finally, the polarization deduced from Eq.~(\ref{sigmaint}) is
directed parallel to $\bm N$
\begin{subequations}
\begin{equation}\label{t+ aggr}
    \bm t_+\parallel\bm N,
\end{equation}
and its degree equals
\begin{equation}\label{Paggr}
\mathtt{P}_{\text{net}}=\frac{N^2}2\frac1{1+\frac{3x_\omega^2}{4(1-x_\omega)}}.
\end{equation}
\end{subequations}
The semi-classical limit ($x_\omega\to0$) of (\ref{Paggr}) at $N=1$
agrees with the polarization $\frac12$ of dipole radiation from a
classical charged particle in a planar undulator \cite{BKStrakh}.
The $x_\omega$-dependent factor describes the polarization
suppression due to the photon recoil. The function
$\mathtt{P}_{\text{net}}/N^2$ is shown in Fig.~\ref{fig:Net} by
solid curve.

The practical value of the non-zero net polarization is that once
there is a target with sizeable $N^2/2$, it suffices for obtaining a
polarized gamma-ray beam without the need for narrow collimation and
particular target thinness \cite{Uberall}. The common known drawback
of incoherent bremsstrahlung radiation is its continuous spectrum,
but that may be overcome by measurement of the energies of all final
products of the induced reactions.

Vice versa, measurement of the net polarization may be used as a
technique for empirical determination of $\bm N$ for a given target.
If one's aim is to obtain a polarized photon beam, this will be
equivalent to the source calibration \emph{in situ}.

\subsection{Angular distributions}

\subsubsection{Unpolarized intensity}

\begin{figure}
\includegraphics{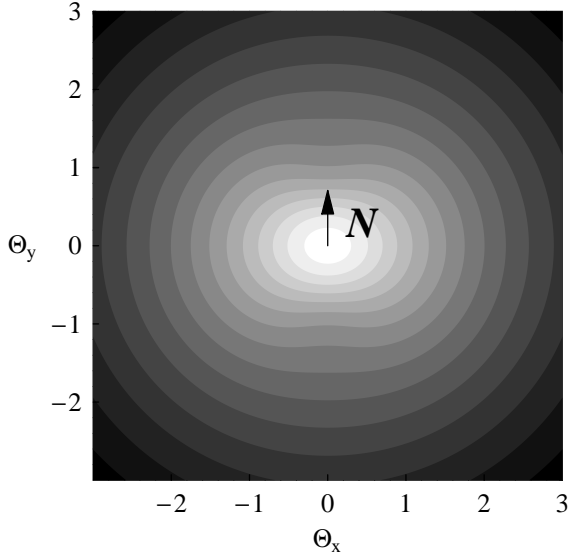}
\caption{\label{fig:dSigmaColor} Logarithm of the unpolarized
differential cross-section, as a function of $\bm\Theta$ (radiation
angles in units $\gamma^{-1}$), at $\bm N$ independent of
$\bm\Theta$, for $N^2=\frac12$, $x_\omega=\frac13$. }
\end{figure}

Inspection of Eq.~(\ref{sigmaunpol}) reveals that the azimuthal
anisotropy embodied by the quadrupole dependence on the angle
$\phi_{\bm\Theta}$ between $\bm\Theta$ and $\bm N$,
\begin{equation*}
    N^2\Theta^2-2(\bm N\cdot\bm\Theta)^2=-N^2\Theta^2\cos2\phi_{\bm\Theta},
\end{equation*}
is sizeable only when $N^2\sim1$, and only at angles $\Theta\sim1$.
At $\Theta\ll1$, or $\Theta\gg1$, the \emph{unpolarized} radiation
differential intensity isotropizes and becomes independent of $N^2$
at all (however, the polarization will not be neither isotropic, nor
small there -- see Eq.~(\ref{Pasymp}) below). The distribution of
unpolarized intensity (\ref{sigmaunpol}) in the $\bm\Theta$ plane,
for constant $\bm N$, at examplary values $N^2=\frac12$ and
$x_\omega=\frac13$, is illustrated in Fig.~\ref{fig:dSigmaColor}. As
compared with Fig.~\ref{fig:UnpolInt}, no dips are left at $N^2$
that small (and $x_\omega$ that large), but there still remains a
noticeable azimuthal anisotropy, the radiation intensity being
enhanced in a bar \emph{orthogonal} to $\bm N$, because dipole
radiation intensity is known to be largest in directions orthogonal
to that of the acceleration (the much greater deflection due to the
transverse recoil from the photon emission proves to be of no
consequence there). In contrast, electron multiple scattering
diffusion in the sample will be fastest in direction \emph{parallel}
to $\bm N$.

If resolution of radiation angles is feasible in the experiment,
measurement of the radiation azimuthal anisotropy (say, at
$\Theta\approx1$) may offer a method for parameter $\bm N$
determination for a given target. Other methods are based on
polarization measurements, needing no electron detection. We now
proceed with discussion of the polarization angular distribution.

\subsubsection{Polarization}\label{subsec:PolAngDistr}

As long as linear polarization is a vector quantity, to handle it
practically, it is best to know its absolute magnitude (degree) and
the direction. However, those variables are not in a linear relation
to the calculated differential probability, which, after momentum
averaging in matter, turns to a generic kind of tensor in $\bm e'$
-- see (\ref{sigmarad2}). This may prompt one to deal with
polarization asymmetries in some fixed coordinate frame, such as
Stokes parameters. But actually, in 2 transverse dimensions
expressing the polarization direction explicitly is not difficult at
all, involving at the most quadratic equations.

From Eq.~(\ref{2Tij-Tdeltaij}), the polarization degree is extracted
as an asymmetry
\begin{eqnarray}\label{Pgeneral}
\mathtt{P}(\bm\Theta ,x_\omega;\bm N)&\overset{\text{def}}=&\frac{\underset{\bm e'}\max\left\langle \frac{dW_{\text{dip}}}{d\Gamma_{k}}\right\rangle-\underset{\bm e'}\min\left\langle \frac{dW_{\text{dip}}}{d\Gamma_{k}}\right\rangle}{\underset{\bm e'}\max\left\langle \frac{dW_{\text{dip}}}{d\Gamma_{k}}\right\rangle+\underset{\bm e'}\min\left\langle \frac{dW_{\text{dip}}}{d\Gamma_{k}}\right\rangle}\nonumber\\
&=&\frac{\alpha}{2\pi^2}\frac{\left\langle\bm
    q^2_\perp\right\rangle_{\text{rad}}(1-x_\omega)}{m^2\left(
1+\Theta^2\right)^2}\frac{2\lambda_+[T_{ij}]}{x_\omega\left\langle\frac{dW_{\text{unpol}}}{dx_\omega d^2\Theta}\right\rangle}\quad\nonumber\\
&=&\frac{\lambda_+[T_{ij}]}{\frac{x_\omega^2}{2(1-x_\omega)}+\frac{1+\Theta^4+2\left[N^2\Theta^2-2(\bm
N\cdot\bm\Theta)^2 \right]}{\left( 1+\Theta^2\right)^2}},
\end{eqnarray}
where $\lambda_+[T_{ij}]$ is the positive eigenvalue of tensor
$\left(2T_{ij}-T\delta_{ij} \right)$. In terms of the latter,
$\lambda_+$ may be expressed as
\begin{equation}\label{la}
\lambda_+=\sqrt{\frac{1}{2}\left(2T_{ij}-T\delta_{ij}
\right)\left(2T_{ij}-T\delta_{ij}
\right)}\equiv\sqrt{2T_{ij}T_{ij}-T^2} .
\end{equation}
In our case (\ref{Tij def}), tensor $T_{ij}$ is formed by two
vectors:
\begin{equation}\label{T=aa-bb}
T_{ij}=a_ia_j-b_ib_j
\end{equation}
with
\begin{equation}\label{a}
a_i=G_{im}N_m,\qquad b_i=\frac{\sqrt{2(1-N^2)}}{1+\Theta^2}\Theta_i
\end{equation}
(those vectors are not mutually orthogonal, in general).
Substituting Eq.~(\ref{T=aa-bb}) to Eq.~(\ref{la}), one
straightforwardly evaluates
\begin{subequations}
\begin{eqnarray}
\lambda_+&=&\sqrt{(\bm a^2-\bm b^2)^2+4[\bm a \times \bm
b]^2}\label{()^2+[]^2}\label{sqrt}\\
&\equiv&\left|\bm a -\bm b\right|\left|\bm a +\bm
b\right|\qquad\label{modmod}
\end{eqnarray}
\end{subequations}
Eigenvectors of a tensor of the form (\ref{T=aa-bb}) can be
expressed covariantly in terms of the vectors $\bm a$, $\bm b$:
\begin{subequations}
\begin{eqnarray}
\bm t_{\pm}&\parallel&2(\bm a\cdot\bm b)\bm a-(\bm a^2+\bm b^2)\bm b\pm\lambda_+\bm b\label{tpm}\\
&\parallel&2(\bm a\cdot\bm b)\bm b-(\bm a^2+\bm b^2)\bm
a\mp\lambda_+\bm a,\label{tmp}
\end{eqnarray}
\begin{equation}\label{t+ perp t-}
    \bm t_+\perp\bm t_-
\end{equation}
\end{subequations}
(the coefficients at $\bm a$, $\bm b$ in (\ref{tpm}-\ref{tmp}) are
found by solving a system of two linear equations).

Substitution of (\ref{a}) into Eqs.~(\ref{sqrt}, \ref{tpm}) leads to
representations
\begin{widetext}
\begin{eqnarray}
\lambda_+&=&\sqrt{\left[N^2-\frac{4(\bm
N\cdot\bm\Theta)^2+2(1-N^2)\Theta^2}{\left( 1+\Theta^2\right)^2}
\right]^2+\frac{8(1-N^2)[\bm\Theta \times \bm N]^2}{\left(
1+\Theta^2\right)^2}},\label{lambda+}\\
%&\\ \leq&N^2+2\frac{(1-N^2)\Theta^2-2(\bm N\cdot\bm\Theta)^2}{\left(1+\Theta^2\right)^2},\label{lambda+bound}
\bm t_\pm&\parallel&2\frac{1-\Theta^2}{1+\Theta^2}(\bm
N\cdot\bm\Theta)\bm N+\left[\frac{2\left[2(\bm
N\cdot\bm\Theta)^2+N^2-1\right]\Theta^2}{\left(
1+\Theta^2\right)^2}-N^2\pm\lambda_+ \right]\bm\Theta.\label{t+}
% \\ &\parallel&-4(1-N^2)\frac{1-\Theta^2}{(1+\Theta^2)^3}(\bm N\bm\Theta)\bm\Theta\nonumber\\&-&\left[N^2-\frac{4(\bm N\bm\Theta)^2}{(1+\Theta^2)^2}+\frac{2(1-N^2)}{(1+\Theta^2)^2}\Theta^2\pm\lambda_\pm\right]\left(-\bm N+\frac{2(\bm N\bm\Theta)}{1+\Theta^2}\bm\Theta\right)
\end{eqnarray}
\end{widetext}

One may note that at $\Theta\gg1/N$ the polarization picture
simplifies, tending to
\begin{equation}\label{Pasymp}
    \mathtt{P}\to N^2\mathtt{P}_{\text{max}}(x_\omega),\quad \bm t_+\parallel\bm N-\frac{2(\bm
    N\cdot\bm\Theta)}{\Theta^2}\bm\Theta. \quad (\Theta\gg1/N)
\end{equation}
So, the polarization distribution shape in this region is the same
as in extremely anisotropic case $|\bm N|=1$, or for non-averaged
$x_\omega\frac{dW_{\text{rad}}}{dx_\omega d^2\Theta}$ at definite
$\bm q_\perp$ (see Sec.~\ref{sec:fact}, Fig.~\ref{fig:Graph1}),
except that the polarization degree is $\propto N^2$
\footnote{However, for $\Theta$ large enough $N$ should become
$\Theta$-dependent (at least, model (\ref{rowN2}) suggests so), and
then polarization at large angles will vanish.}. However at
practice, since $N$ decreases with increasing $\Theta$ (according to
Eq.~(\ref{rowN2}), $N\underset{\Theta\to\infty}\sim\Theta^{-2}$), at
sufficiently large $\Theta$ isotropy must set in. We refrain from
studying the transition to this regime insofar as it is model- and
process- dependent.

\begin{figure}
\includegraphics[width=\columnwidth]{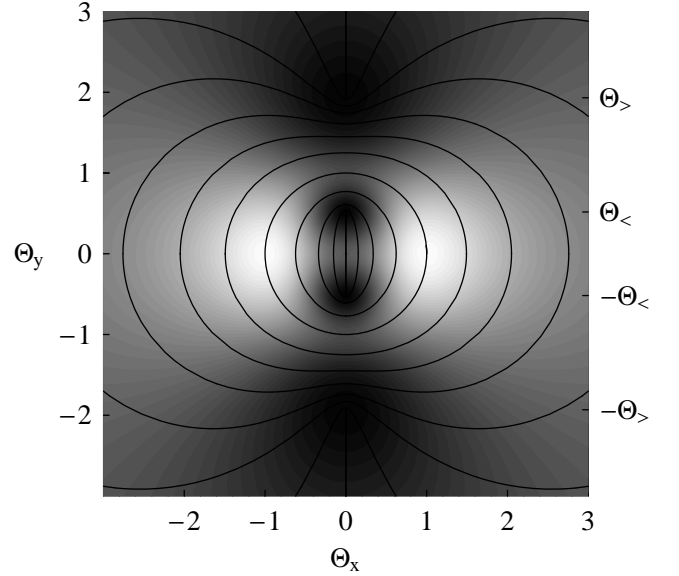}
\caption{\label{fig:PContourPlotN=0.5} Angular distribution of the
polarization degree (density plot) and orientation (black curves) at
$\bm N$ independent of $\bm\Theta$, for $N^2=\frac12$,
$x_\omega=\frac13$. For principal profiles of this distribution --
see Figs.~\ref{fig:PparallN,N2=0.5}(a), \ref{fig:PperpN,N2=0.5}(b)
below, middle curves. The direction of $\bm N$ is vertical. At
$|\bm\Theta|\to\infty$ polarization degree tends to a constant value
$N^2$. The unpolarized radiation intensity for the same parameters
is displayed in Fig.~\ref{fig:dSigmaColor}.}
\end{figure}

The complete polarization picture (the polarization degree and
direction), is shown in Fig.~\ref{fig:PContourPlotN=0.5}. It appears
to have a richer structure than the corresponding unpolarized
intensity in Fig.~\ref{fig:dSigmaColor}. A novel feature at $N<1$ is
that the knot points (\ref{Theta plus-minus}) split, admitting the
``polarization flow" into the gaps. Polarization zero positions now
can be found from setting $\mathtt{P}$, or $\lambda_+$ as given by
Eq.~(\ref{lambda+}), equal to zero. That yields
\begin{eqnarray}\label{Theta><}
    \bm\Theta_>&=&\pm\frac{\bm
    N}{N^2}\sqrt{1+\sqrt{1-N^4}},\\
    \bm\Theta_<&=&\pm\frac{\bm
    N}{N^2}\sqrt{1-\sqrt{1-N^4}}.
\end{eqnarray}
The mean geometric value of distances to these points from the
origin equals 1:
\begin{equation}\label{mean geom}
    \sqrt{\Theta_>\Theta_<}=1\
\end{equation}
(this can be traced to the fact that on the sphere of radiation
directions in the initial electron rest frame those points are
located symmetrically relative to plane $z=0$, and upon the
stereographic projection they become conjugate with respect to the
unit circle, see Sec.~\ref{subsec:stereogr}, Theorem 1), whereas the
gap width
\begin{equation}\label{clearance}
    \Theta_>-\Theta_<=\frac{\sqrt2\sqrt{1-N^2}}N
\end{equation}
exhibits ``threshold behavior" as $N$ departs from 1. However, now
the points of zero polarization do not correspond to any dips in the
radiation intensity (cf. Fig.~\ref{fig:dSigmaColor}).

Determination of the bremsstrahlung polarization zeroes may serve as
another calibration method for the parameter $\bm N$. The virtue of
this method is that it does not require absolute measurements of
intensity, but of the angles only. When $N$ is small, it is
convenient to measure it through measurement of $\Theta_<$, which is
$\Theta_<\sim N$, in contrast to $\mathtt{P}_{\text{net}}\sim N^2$.
Vice versa, when $N^2$ is close to 1 (say, in coherent
bremsstrahlung), it is convenient to measure (\ref{clearance}) (due
to the square root dependence).

The analytic form of the polarization tangential curves in general
case is complicated (though, stereographic projection can offer some
simplifications), and we do not contemplate determining it here. At
least, at $\Theta=1$ it is apparent that $\bm
t_-\parallel\bm\Theta$, hence $\bm t_+\perp\bm\Theta$, i. e.
\footnote{For the ``+" sign, the r. h. s. of (\ref{t+}) vanishes,
giving indeterminancy for $\bm t_+$. Fortunately, for $\bm t_-$ such
a problem does not arise.}, polarization direction is steered along
the unit circle, anyway.

\begin{figure}
%\begin{subfigures}
\includegraphics{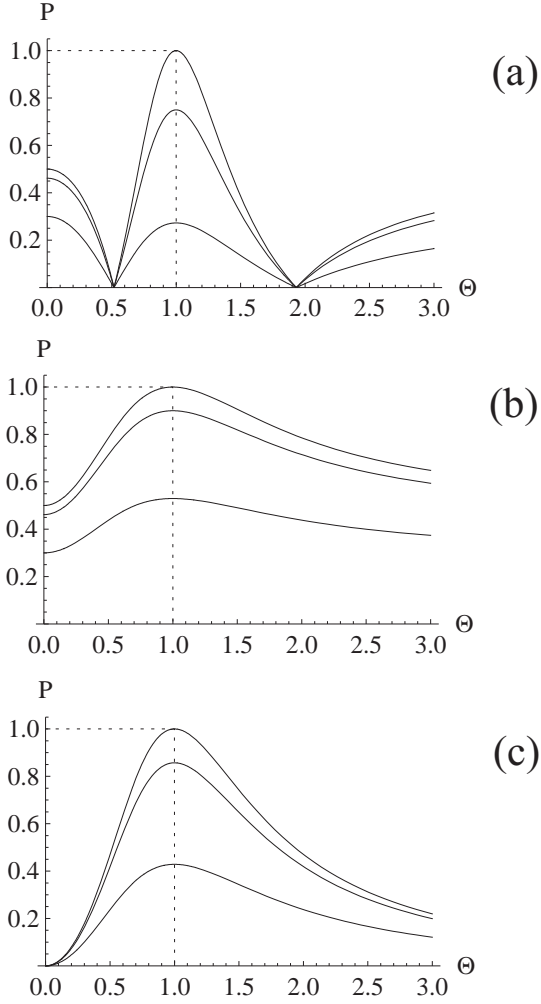}
%\end{subfigures}
\caption{ (a).\label{fig:PparallN,N2=0.5} Polarization degree at
$\bm\Theta\parallel\bm N$ for $\bm N$ independent of $\omega$,
$\bm\Theta$, having magnitude $N^2=\frac12$, and for photon energy
fractions $x_\omega=0,\frac13$, and $\frac23$ (top to bottom).
Orientation of the polarization depends on the interval -- see
Eqs.~(\ref{intrls}). (b).\label{fig:PperpN,N2=0.5} Same as (a),
except that profile $\bm\Theta\perp\bm N$ is shown. Polarization
orientation is $\parallel\bm N$. (c).\label{fig:Pol-centr}
Polarization of bremsstrahlung on an isotropic target ($N=0$) for
same values of $x_{\omega}$.}
\end{figure}

To gain more quantitative understanding of the profile of
polarization distribution, it is instructive to examine its two
principal profiles: $\bm\Theta\parallel\bm N$ and $\bm\Theta\perp\bm
N$. In those cases, $\bm a\parallel \bm b$, or $\bm a\perp \bm b$,
owing to which Eqs. (\ref{lambda+}, \ref{t+}) substantially
simplify.

\noindent (i) If $\bm\Theta\parallel\bm N$,
\begin{equation}\label{lambda||}
    \lambda_+=\left|N^2-\frac{2\Theta^2(1+N^2)}{(1+\Theta^2)^2}\right|,
\end{equation}
\begin{subequations}\label{intrls}
\begin{eqnarray}
\bm t_+\parallel\bm N&\,& \text{if either}\quad
|\bm\Theta|<\Theta_<,\,\,\,\text{or}\,\,\,|\bm\Theta|>\Theta_>,\qquad\label{2 intervals}\\
\bm t_+\perp\bm N &\,&\text{in the
interval}\quad\Theta_<<|\bm\Theta|<\Theta_>\label{1 interval}
\end{eqnarray}
\end{subequations}
(Eq.~(\ref{2 intervals}) is inferred from Eq.~(\ref{t+}) with the
upper sign, Eq.~(\ref{1 interval}) -- from Eq.~(\ref{t+}) with the
lower sign). The polarization degree (\ref{Pgeneral}) through
(\ref{lambda||}) reduces to
\begin{equation}
\mathtt{P}=\frac{\left| N^2\left(
1+\Theta^4\right)-2\Theta^2\right|}{1+\Theta^4-2N^2\Theta^2+\frac{x_\omega^2}{2(1-x_\omega)}\left(
1+\Theta^2\right)^2}.
\end{equation}
This function is displayed in Fig.~\ref{fig:PparallN,N2=0.5}(a). It
drops to zero at $\Theta=\Theta_{+<},\Theta_{+>}$, has maxima at
$\Theta=0$ and $\Theta=\infty$, where it achieves the same values
\begin{equation}\label{}
    \mathtt{P}(\bm\Theta=0)=\mathtt{P}(\bm\Theta=\infty)=N^2\mathtt{P}_{\text{max}}(x_\omega)=\frac{2N^2\left(1-x_\omega\right)}{1+\left(1-x_\omega\right)^2},
\end{equation}
and at $\Theta^2=1$, where
\begin{equation}\label{}
    \mathtt{P}(\bm\Theta=\pm\bm N/|\bm
N|)=\frac1{1+\frac{x_\omega^2}{(1-x_\omega)(1-N^2)}}.
\end{equation}

\noindent(ii) In the case of orthogonal profile $\bm\Theta\perp\bm
N$,
\[
\bm t_+\perp\bm t_-\parallel\bm\Theta
\]
(inferred from Eq.~(\ref{t+}) with the lower sign),
Eq.~(\ref{lambda+}) turns to
\[
\lambda_+=N^2+(1-N^2)\frac{2\Theta^2}{\left(1+\Theta^2\right)^2},
\]
and so polarization degree (\ref{Pgeneral}) becomes
\begin{equation}
\mathtt{P}=\frac{N^2\left(
1+\Theta^4\right)+2\Theta^2}{1+\Theta^4+2N^2{\Theta}^2+\frac{x_\omega^2}{2(1-x_\omega)}\left(
1+\Theta^2\right)^2}
\end{equation}
(shown in Fig.~\ref{fig:PperpN,N2=0.5}(b)). It reaches a maximum at
$\Theta^2=1$, where
\begin{eqnarray}\label{PmaxN}
\mathtt{P}_{\text{max}}(x_\omega,N^2)&=&\frac{1}{1+\frac{x_\omega^2}{(1-x_\omega)(1+N^2)}}\\
&\leq&\mathtt{P}_{\text{max}}(x_\omega,1)\equiv\mathtt{P}_{\text{max}}(x_\omega).\nonumber
\end{eqnarray}
(As Fig.~\ref{fig:PContourPlotN=0.5} indicates, and can be proven
based on Eqs.~(\ref{Pgeneral}, \ref{lambda+}), this is the absolute
maximum for all $\bm\Theta$). At $\Theta=0$ and $\infty$,
polarization is minimal with the value
$N^2\mathtt{P}_{\text{max}}(x_\omega)$.

There is a feature already mentioned in Sec.~\ref{subsec:stereogr},
that at $\Theta=1$ the polarization is capable of achieving 1, in
spite that we are summing portions of completely polarized light,
but with different polarization orientations (vector $\mathsf
G\hat{\bm q}_\perp$ with $\mathsf G$ given by Eq.~(\ref{Gim}),
generally, rotates along with $\hat{\bm q}_\perp$). This is explaned
by recalling the pattern of polarization alignment at a given
$\hat{\bm q}_\perp$ (Fig.~\ref{fig:Graph1}): since at $\Theta=1$
polarization is oriented along a perfect circle centered at the
origin of the plane, and that circle remains self-collinear under
the rotations of $\hat{\bm q}_\perp$ corresponding to azimuthal
averaging, the absolute polarization at this special angle is
unaffected by the scattering isotropization.

\begin{figure}
\includegraphics{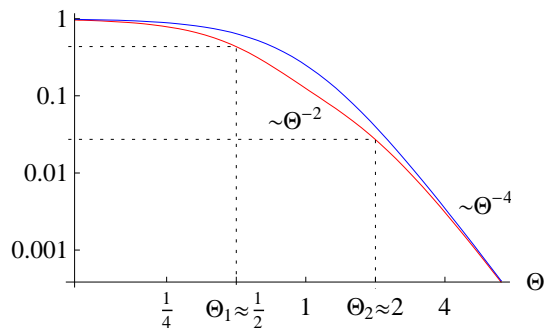}
\caption{\label{fig:LogLogPlot3} Log-log plot of the dipole
bremsstrahlung angular distribution in an isotropic target. Lower
(red) curve -- semi-classical limit
$J_{\mathrm{BH}}\left(0,\Theta\right)/J_{\mathrm{BH}}\left(0,0\right)$.
Upper (blue) curve -- hard limit
$J_{\mathrm{BH}}\left(1,\Theta\right)/J_{\mathrm{BH}}\left(1,0\right)$,
equal to the differential intensity pre-factor $1/(1+\Theta^2)^2$.
The soft radiation angular distribution exhibits two ``knees", at
$\Theta\approx\frac12$ and $\Theta\approx2$ whereas the hard
radiation only has one, at $\Theta\approx1$.}
\end{figure}

To conclude this subsection, note that the spots of high and
directionally stable polarization at
$(\Theta_x,\Theta_y)\approx(\pm1,0)$ are also attractive for
extraction of a polarized photon beam by the radiation collimation
technology. However, with a collimation facility at disposal, one
can obtain a polarized photon beam on an isotropic target as well --
see the next subsection (and \cite{Taylor-Mozley}). One should mind
also that at $\Theta\approx1$ the radiation intensity is by an order
of magnitude lower than at $\Theta\approx0$ (see
Fig.~\ref{fig:LogLogPlot3}). On the other hand, the region
$\Theta\approx0$ is polarized, too, though $2/N^2$ times weaker. But
the latter drawback may be compensated by an order-of-magnitude
higher intensity. Thus, for extraction of a polarized photon beam
one may crop an angular strip at $|\Theta_y|\lesssim0.7$.

\subsubsection{Isotropic target ($N=0$)}\label{sec:spher}
Since most substances of natural origin are fairly isotropic on
macro-scales, all early studies of bremsstrahlung presumed the
scattering isotropy. The results of classic works \cite{May Wick}
are readily reproduced from our generic equations.

Setting in our Eq.~(\ref{sigmarad}) $\bm N=0$, one reproduces the
equation for the polarization-dependent differential cross-section
of bremsstrahlung in an isotropic medium obtained by May and Wick
\cite{May Wick}. To obtain separately the corresponding unpolarized
differential cross-section and polarization, it suffices to let $\bm
N=0$ in our Eqs.~(\ref{sigmaunpol}, \ref{lambda+}, \ref{t+}):
\begin{eqnarray}\label{sigmacsymm}
    J_{\mathrm{BH}}\left(x_\omega,\Theta\right)=x_\omega\left\langle\frac{dW_{\text{isotr}}}{dx_\omega d^2\Theta}\right\rangle=x_\omega\left\langle\frac{dW_{\text{unpol}}}{dx_\omega d^2\Theta}\right\rangle\bigg\rvert_{\bm
    N=0}\nonumber\\
    =\frac{\alpha}{2\pi^2}\frac{\langle\bm
    q_\perp^2\rangle_{\text{rad}}}{m^2\left(1+\Theta^2\right)^2}\left\{2(1-x_\omega)\frac{1+\Theta^4}{\left(1+\Theta^2\right)^2}+x_\omega^2\right\},\quad
\end{eqnarray}
\begin{eqnarray}\label{Pcsymm}
\mathtt{P}_{\text{isotr}}(\Theta,x_\omega)&=&\mathtt{P}(\Theta,x_\omega)\big\rvert_{\bm
N=0}\nonumber\\
&=&\frac{2\Theta^2}{1+\Theta^4+\frac{x_{\omega}^2}{2(1-x_{\omega})}\left(1+\Theta^2\right)^2},\quad
\end{eqnarray}
\begin{equation}\label{t+ perp Theta}
    \bm t_+\perp\bm\Theta.
\end{equation}
Eq.~(\ref{sigmacsymm}) may be regarded as Bethe-Heitler's radiation
intensity angular distribution in the leading logarithmic
approximation (the logarithmic factor being contained in $\langle\bm
q_\perp^2\rangle_\text{rad}$). Relation (\ref{t+ perp Theta}) is the
observation of May and Wick \cite{May Wick}. Its interpretation is
that dipole emissivity dominates in directions orthogonal to that of
the acceleration. I. e., the sample of events containing a photon at
an angle $\bm\Theta$ is biased towards momentum transfers in matter
orthogonal to $\bm\Theta$; these events are then likely to contain
photon polarization collinear with $\bm q$, and thus perpendicular
to $\bm\Theta$.

Concerning the above angular distribution shapes, we may add two
remarks.

\paragraph{``Double-knee" in the angular distribution of soft radiation.} The function
\begin{equation}\label{func}
    \frac{1+\Theta^4}{\left(1+\Theta^2\right)^2}\equiv1-\frac{2\Theta^2}{\left(1+\Theta^2\right)^2},
\end{equation}
determining at small $x_\omega$ the r.h.s. of
Eq.~(\ref{sigmacsymm}), has a minimum at $\Theta=1$ (precisely where
the dips in the non-averaged $dW_{\text{dip}}/d\Gamma_{k}$ are
located, see Fig.~\ref{fig:UnpolInt}). However, function
$J_{\mathrm{BH}}\left(0,\Theta\right)$, in fact, does not develop
any minimum or shoulder about this point, because it involves yet a
pre-factor $1/(1+\Theta^2)^2$ decreasing steeper than (\ref{func})
rises after its minimum. Nevertheless, some imprint of function
(\ref{func}) remains in behavior of the Bethe-Heitler radiation
angular distribution. To show that, let us plot logarithm of
$J_{\mathrm{BH}}\left(0,\Theta\right)$ vs. the logarithm of $\Theta$
($x_\omega\approx0$ is taken to enhance the relative contribution of
(\ref{func})).
%\footnote{It may
%be mentioned that in a real medium at frequency fractions as small
%as $x_\omega\leq10^{-4}$ there may exists yet longitudinal density
%(Ter-Mikaelian) effect \cite{Ter-Mik}, leading to the radiation
%suppression and broadening. However, the dipole approximation
%condition (\ref{xto0}) will break down earlier, so we can not meet
%this effect here.})  as a function of $\Theta$, using double
%logarithmic scale (see Fig.~\ref{fig:LogLogPlot3}).
In such kind of a plot linear dependence corresponds to a power-law
falloff. As compared with the behavior of the pre-factor
$1/(1+\Theta^2)^2$, which has on this plot only one ``knee" at
$\Theta\approx1$, $J_{\mathrm{BH}}(0,\Theta)$ apparently has two
``knees" -- at about $\Theta_1\approx0.5$ and at $\Theta_2\approx2$.
In between of those two ``knees" the behavior is close to
$\sim\Theta^{-2}$. Beyond $\Theta_2$, the falloff power turns to
$\sim\Theta^{-4}$, as is required by the quasi-Rutherford law
mentioned in Sec.~\ref{subsec:lab}. This 2-knee shape of the
radiation angular distribution may be worth minding at poor
statistics measurements, since at the second knee the differential
cross-section is already down by a factor of nearly $10^{-2}$. But
as $x_\omega$ grows, the distribution approaches the 1-knee limiting
form.

\paragraph{Polarization maximum and its non-dipole destruction.}
For what concerns polarization (shown in
Fig.~\ref{fig:Pol-centr}(c)), again, the prominent feature is that
it reaches 100\% at $x_{\omega}\ll1$, $\Theta=1$, for the reasons
already explained (but the example at $\bm N=0$ is just the most
spectacular). However, with the account of non-dipole effects (see,
e. g., \cite{May}), polarization in the region $\Theta\approx1$ must
deplete, because, at a definite but large $\hat{\bm q}_\perp$ the
polarization tangential curves are still circles, but their centres
are shifted by a vector $\frac{\bm q_\perp}{2m}$, and none of them
coincides with the origin anymore.

\section{Summary}\label{sec:summary}

The present study suggests that there must exist macroscopic
targets, on which relativistic electron scattering, and hence the
accompanying forward radiation, possesses a high degree of azimuthal
anisotropy. Suitable examples are single crystals oriented by one of
their strong crystallographic axes at moderately small orientation
angles $\sim10^{-1}\div10^{-2}\text{rad}$ with respect to the
electron beam direction. The azimuthal anisotropy of the scattering
is quantified by parameter $\bm N$ introduced in
Sec.~\ref{sec:integr}. As we have investigated, the azimuthal
anisotropy is partially spoiled by nuclei vicinities and by the
radiation recoil (see Eq.~(\ref{rowN2})), but nonetheless, values
$N^2\geq0.5$ look realistic.

A straightforward application of polarized bremsstrahlung is for
preparation of polarized photon beams (of continuous spectrum).
Actually, there is a number of options for extracting polarized
photons from the bremsstrahlung flux:

\noindent (i) If only isotropic targets are at disposal, there is no
alternative to the traditional method \cite{Taylor-Mozley} of
collimating the bremsstrahlung photon flux around the angle
$\Theta=1$, i. e., $\theta=1\cdot\gamma^{-1}$.

\noindent (ii) If the use of absorber collimators is prohibitive due
to radiation angle smallness, as it tends to be at $\gamma>10^4$,
one needs an intrinsically anisotropic target, the aggregate
(naturally narrow) cone of photons emitted on which is polarized.
But its polarization degree, according to Eq.~(\ref{Paggr}), is
$\leq N^2/2$, with $N^2<1$. For efficiency of such a polarized beam,
it is desirable to have $N^2$ at least $\sim0.7\div0.8$.

\noindent (iii) Ultimately, if both a collimation tool and an
intrinsically  anisotropic target are available, one may either look
for the highest polarization degree, isolating one of the two spots
of enhanced polarization (see Fig.~\ref{fig:PContourPlotN=0.5}). Or,
if moderate polarization degree is acceptable provided the beam
intensity is high, there is an option of collimating out the strip
of angles perpendicular to $\bm N$, in between of the polarization
zeroes.

Another application of polarized bremsstrahlung from relativistic
electrons is for measuring intrinsic anisotropy of the medium the
electrons move in. Such kind of diagnostics may be useful during
various experiments on relativistic electron interaction with
crystal-based targets. Obviously, the radiation leaving the target
without much rescattering is better suited for detection than the
emitting electrons themselves. Again, for measurement of $\bm N$ by
the bremsstrahlung yield one can employ a number of techniques:

\noindent  (i) If the radiation angles can not be resolved, one has
to measure $\mathtt{P}_{\text{net}}\propto\frac{N^2}2$
(Eq.~(\ref{Paggr}), Fig.~\ref{fig:Net}).

\noindent (ii) If the radiation angles are resolvable, one can
estimate $\bm N$ by the bremsstrahlung intensity azimuthal
anisotropy, not employing the polarization detection
(Eq.~(\ref{sigmaunpol}), Fig.~\ref{fig:UnpolInt}).

\noindent(iii) For a finest measurement of $N$, particularly under
conditions when it is close to 0 or 1, one can use the method of
finding polarization zero locations in the angular distribution
(Fig.~\ref{fig:PContourPlotN=0.5}). For $N$ small, it is convenient
to measure $\Theta_<$, since it is linear in $N$, whereas for
$N\to1$ -- the gap $\Theta_>-\Theta_<$, which is $\sim\sqrt{1-N^2}$
(see Sec.~\ref{subsec:PolAngDistr}).

In conclusion, let us yet draw attention to a useful methodic notion
-- the stereographic projection relation between the laboratory
frame and the initial electron rest frame
(Sec.~\ref{subsec:stereogr}). It helps revealing various symmetry
properties of relativistic particle bremsstrahlung angular
distributions, and may facilitate calculation of bremsstrahlung
characteristics in some cases. It must also survive in the
non-dipole bremsstrahlung case, which we hope to investigate
elsewhere.

\vspace{3mm}

\thanks{ }
\textbf{Acknowledgements}

The author wishes to thank N.~P.~Merenkov and A.~V.~Shchagin for
useful discussions.

%--------------------   Bibliography  ------------------------------------%

%Just because of unusual number of tables stacked at end
\bibliography{RefsBremss.bib}% Produces the bibliography via BibTeX.

\begin{thebibliography}{99}
\bibitem{Proc gamma-nucl} \emph{Proceedings of Int. Workshop on Physics with
GeV electrons and gamma-rays, 2001}, edited by T.~Tamae \emph{et
al.} (Tokyo, Universal Acad. Press, 2001) (Also in: Frontiers
science series. No. 36).
\bibitem{RHESSI} M.~L.~McConnell, J.~M.~Ryan, D.~M.~Smith, R.~P.~Lin, and
A.~G.~Emslie, arXiv:astro-ph/0209384.
\bibitem{nucl polarimeters}
F.~Rambo \emph{et al.}, Phys. Rev. C \textbf{58}, 489 (1998);
N.~Cabibbo, Phys. Rev. Lett. \textbf{7}, 386 (1961).
\bibitem{pol-through-or-cryst}
K.~Kirsebom \emph{et al.}, Phys. Lett. B \textbf{459}, 347 (1999);
A.~Apyan \emph{et al.}, Phys. Rev. ST-AB \textbf{11}, 041001 (2008).
\bibitem{Gemmel-Kumakhov}
D.~S.~Gemmel, Rev. Mod. Phys. \textbf{46}, 129 (1974);
V.~V.~Beloshitsky and F.~F.~Komarov. Phys. Rep. \textbf{93}, 117
(1982).
\bibitem{Diambrini}
G. Diambrini Palazzi, Rev. Mod. Phys. \textbf{40}, 611 (1968).
\bibitem{Ter-Mik}
M.~L.~Ter-Mikaelian, \emph{High-Energy Electromagnetic Processes in
Condensed Media} (Wiley-Interscience, N.Y., 1972).
\bibitem{WW}
E.~Fermi, Z. Phys. \textbf{29}, 315 (1924); C.~F.~Weizs\"{a}cker,
Z.~Phys. \textbf{88}, 612 (1933); E.~J.~Williams, Proc. Roy. Soc.
\textbf{139}, 163 (1933); Phys. Rev. \textbf{45}, 729 (1934).
\bibitem{Bertulani}
C.~A. Bertulani and G.~Baur, Phys. Rep. \textbf{161}, 299 (1986).

\bibitem{BKStrakh}
V.~N.~Baier, V.~M.~Katkov, and V.~M.~Strakhovenko,
\emph{Electromagnetic processes at high energies in oriented
crystals} (World Scientific, Singapore, 1998).
\bibitem{Cheng-Wu}
H.~Cheng and T.~T.~Wu. \emph{Expanding Protons: Scattering at High
Energies} (M.I.T. Press, Cambridge, MA, 1987).
\bibitem{Olsen-Maximon-Wergeland} H.~Olsen, L.~C.~Maximon, and
H.~Wergeland, Phys. Rev. \textbf{106}, 27 (1957); V.~N.~Baier and
V.~M.~Katkov, Sov. Phys. JETP \textbf{28}, 807 (1968); \textbf{28},
854 (1968); A.~I.~Akhiezer, V.~F.~Boldyshev, and N.~F.~Shul'ga, Yad.
Fiz. \textbf{22}, 1185 (1975).



\bibitem{BLP}
V.~B.~Berestetskii, E.~M.~Lifshitz, and L.~P.~Pitaevskii,
\emph{Quantum Electrodynamics} (Pergamon-Press, Oxford, 1982).

\bibitem{Klein}
O.~Klein and Y.~Nishina, Z. Phys. \textbf{52}, 853
(1929).

\bibitem{Budnev}
V.~M.~Budnev, I.~F.~Ginzburg, G.~V.~Meledin, and V.~G.~Serbo, Phys.
Rep. \textbf{15}, 181 (1975).

\bibitem{undul}
D.~F.~Alferov, Yu.~A.~Bashmakov, and P.~A.~Cherenkov, Sov. Phys.
Usp. \textbf{32}, 200 (1989).
\bibitem{Rosenfeld-Sergeeva} B.~A.~Rosenfeld and
N.~D.~Sergeeva, \emph{Stereographic projection} (Mir, Moscow, 1977).

\bibitem{Huygens}
G.~Alberi and G.~Goggi, Phys. Rep. \textbf{74}, 1 (1981);
N.~V.~Bondarenko and N.~F.~Shul'ga, Phys. Lett. B \textbf{427}, 114
(1998).
\bibitem{doughnut}
U. Uggerh{\o}j, Rev. Mod. Phys. \textbf{77}, 1131 (2005).

\bibitem{BondCBBC}
M.~V.~Bondarenco, Phys. Rev. A \textbf{81}, 052903 (2010).

\bibitem{Taylor-Mozley}
R.~E.~Taylor and R.~F.~Mozley, Phys. Rev. \textbf{117}, 835
(1960).
\bibitem{Uberall} H. Uberall, Phys. Rev. \textbf{107}, 223
(1957).
\bibitem{May Wick}
M.~M.~May and G.~C.~Wick, Phys. Rev. \textbf{81}, 628 (1951); R.~L.
Gluckstern, M.~H.~Hull, and G.~Breit, Phys. Rev. \textbf{90}, 1026
(1953); H.~Olsen and L.~C.~Maximon, Phys. Rev. \textbf{114}, 887
(1959).
\bibitem{May}
M.~M.~May, Phys. Rev. \textbf{84}, 265 (1951).
\end{thebibliography}

\end{document}